\documentclass[twocolumn,tighten]{aastex631}
\usepackage{CJKutf8}
\usepackage{amsmath}
\usepackage{enumitem}
\usepackage{booktabs,pbox}
\usepackage{bm}

\graphicspath{{./figures/}{../figures/}}

\newcommand{\sbullet}{\mathbin{\vcenter{\hbox{\scalebox{0.7}{$\bullet$}}}}}
\newcommand{\sbsc}[1]{_\mathrm{#1}}

\newcommand{\HI}{\ion{H}{1}}

\newcommand{\CO}[2]{\mbox{$\mathrm{CO}\,(#1\text{--}#2)$}}

\newcommand{\SigSFR}{\Sigma\sbsc{SFR}}
\newcommand{\Siggas}{\Sigma\sbsc{gas}}
\newcommand{\Sigatom}{\Sigma\sbsc{atom}}
\newcommand{\Sigmol}{\Sigma\sbsc{mol}}
\newcommand{\tdep}{t\sbsc{dep}}
\newcommand{\torb}{t\sbsc{orb}}
\newcommand{\SFEorb}{\epsilon\sbsc{orb}}
\newcommand{\tff}{\bar{t}\sbsc{ff}}
\newcommand{\tffsub}[1]{\bar{t}\sbsc{ff,\,#1}}
\newcommand{\SFEff}{\epsilon\sbsc{ff}}

\newcommand{\PDE}{P\sbsc{DE}}
\newcommand{\Yfb}{\Upsilon\sbsc{fb}}

\newcommand{\alphaCO}{\alpha\sbsc{CO}}

\newcommand{\ualphaCO}{\mbox{$\rm M_\odot~pc^{-2}\ (K~km~s^{-1})^{-1}$}}
\newcommand{\uV}{\mbox{$\rm km~s^{-1}$}}

\newcommand{\uSig}{\mbox{$\rm M_\odot~pc^{-2}$}}

\newcommand{\uSigSFR}{\mbox{$\rm M_\odot~yr^{-1}~kpc^{-2}$}}
\newcommand{\uP}{\mbox{$\rm K~cm^{-3}$}}

\shorttitle{PHANGS SF Laws \& Efficiencies}
\shortauthors{Sun et al.}

\begin{document}

\title{Star Formation Laws and Efficiencies across 80 Nearby Galaxies}

\suppressAffiliations

\newcommand{\McMaster}{\affiliation{Department of Physics and Astronomy, McMaster University, 1280 Main Street West, Hamilton, ON L8S 4M1, Canada}}

\newcommand{\CITA}{\affiliation{Canadian Institute for Theoretical Astrophysics (CITA), University of Toronto, 60 St George Street, Toronto, ON M5S 3H8, Canada}}

\newcommand{\OSU}{\affiliation{Department of Astronomy, The Ohio State University, 140 West 18th Avenue, Columbus, OH 43210, USA}}

\newcommand{\CCAPP}{\affiliation{Center for Cosmology and Astroparticle Physics (CCAPP), 191 West Woodruff Avenue, Columbus, OH 43210, USA}}

\newcommand{\Alberta}{\affiliation{Department of Physics, University of Alberta, Edmonton, AB T6G 2E1, Canada}}

\newcommand{\ANU}{\affiliation{Research School of Astronomy and Astrophysics, Australian National University, Canberra, ACT 2611, Australia}}

\newcommand{\ASIAA}{\affiliation{Institute of Astronomy and Astrophysics, Academia Sinica, No. 1, Sec. 4, Roosevelt Road, Taipei 10617, Taiwan}}

\newcommand{\ASTROThreeD}{\affiliation{ARC Centre of Excellence for All Sky Astrophysics in 3 Dimensions (ASTRO 3D), Australia}}

\newcommand{\Bonn}{\affiliation{Argelander-Institut f\"ur Astronomie, Universit\"at Bonn, Auf dem H\"ugel 71, 53121 Bonn, Germany}}

\newcommand{\Carnegie}{\affiliation{Observatories of the Carnegie Institution for Science, 813 Santa Barbara Street, Pasadena, CA 91101, USA}}

\newcommand{\CfA}{\affiliation{Center for Astrophysics $\mid$ Harvard \& Smithsonian, 60 Garden Street, Cambridge, MA 02138, USA}}

\newcommand{\CITEVA}{\affiliation{Centro de Astronomía (CITEVA), Universidad de Antofagasta, Avenida Angamos 601, Antofagasta, Chile}}

\newcommand{\CNRS}{\affiliation{CNRS, IRAP, 9 Av. du Colonel Roche, BP 44346, F-31028 Toulouse cedex 4, France}}

\newcommand{\COOL}{\affiliation{Cosmic Origins Of Life (COOL) Research DAO, coolresearch.io}}

\newcommand{\ESO}{\affiliation{European Southern Observatory, Karl-Schwarzschild Stra{\ss}e 2, D-85748 Garching bei M\"{u}nchen, Germany}}

\newcommand{\Gent}{\affiliation{Sterrenkundig Observatorium, Universiteit Gent, Krijgslaan 281 S9, B-9000 Gent, Belgium}}

\newcommand{\Hawaii}{\affiliation{Institute for Astronomy, University of Hawaii, 2680 Woodlawn Drive, Honolulu, HI 96822, USA}}

\newcommand{\Heidelberg}{\affiliation{Astronomisches Rechen-Institut, Zentrum f\"{u}r Astronomie der Universit\"{a}t Heidelberg, M\"{o}nchhofstra\ss e 12-14, D-69120 Heidelberg, Germany}}

\newcommand{\ICRAR}{\affiliation{International Centre for Radio Astronomy Research, University of Western Australia, 35 Stirling Highway, Crawley, WA 6009, Australia}}

\newcommand{\INAF}{\affiliation{INAF -- Osservatorio Astrofisico di Arcetri, Largo E. Fermi 5, I-50157, Firenze, Italy}}

\newcommand{\IPAC}{\affiliation{Caltech-IPAC, 1200 E. California Blvd. Pasadena, CA 91125, USA}}

\newcommand{\IPARC}{\affiliation{Instituto de F\'{\i}sica de Part\'{\i}culas y del Cosmos IPARCOS, Facultad de Ciencias F\'{\i}sicas, Universidad Complutense de Madrid, E-28040, Spain}}

\newcommand{\IRAM}{\affiliation{Institut de Radioastronomie Millim\'etrique (IRAM), 300 Rue de la Piscine, F-38406 Saint Martin d'H\`eres, France}}

\newcommand{\ITA}{\affiliation{Universit\"{a}t Heidelberg, Zentrum f\"{u}r Astronomie, Institut f\"{u}r Theoretische Astrophysik, Albert-Ueberle-Str 2, D-69120 Heidelberg, Germany}}

\newcommand{\IWR}{\affiliation{Universit\"{a}t Heidelberg, Interdisziplin\"{a}res Zentrum f\"{u}r Wissenschaftliches Rechnen, Im Neuenheimer Feld 205, D-69120 Heidelberg, Germany}}

\newcommand{\JHU}{\affiliation{Department of Physics and Astronomy, The Johns Hopkins University, Baltimore, MD 21218, USA}}

\newcommand{\LAM}{\affiliation{Aix Marseille Univ, CNRS, CNES, LAM (Laboratoire d’Astrophysique de Marseille), Marseille, France}}

\newcommand{\Leiden}{\affiliation{Leiden Observatory, Leiden University, P.O. Box 9513, 2300 RA Leiden, The Netherlands}}

\newcommand{\Liverpool}{\affiliation{Astrophysics Research Institute, Liverpool John Moores University, IC2, Liverpool Science Park, 146 Brownlow Hill, Liverpool L3 5RF, UK}}

\newcommand{\Lyon}{\affiliation{Univ Lyon, Univ Lyon 1, ENS de Lyon, CNRS, Centre de Recherche Astrophysique de Lyon UMR5574, F-69230 Saint-Genis-Laval, France}}

\newcommand{\Maryland}{\affiliation{Department of Astronomy and Joint Space-Science Institute, University of Maryland, College Park, MD 20742, USA}}

\newcommand{\MPE}{\affiliation{Max-Planck-Institut f\"{u}r extraterrestrische Physik, Giessenbachstra{\ss}e 1, D-85748 Garching, Germany}}

\newcommand{\MPIA}{\affiliation{Max-Planck-Institut f\"{u}r Astronomie, K\"{o}nigstuhl 17, D-69117, Heidelberg, Germany}}

\newcommand{\Nagoya}{\affiliation{Department of Physics, Nagoya University, Furo-cho, Chikusa-ku, Nagoya, Aichi 464-8602, Japan}}

\newcommand{\NAOJ}{\affil{National Astronomical Observatory of Japan, 2-21-1 Osawa, Mitaka, Tokyo, 181-8588, Japan}}

\newcommand{\Nichidai}{\affil{Department of Physics, General Studies, College of Engineering, Nihon University, 1 Nakagawara, Tokusada, Tamuramachi, Koriyama, Fukushima, 963-8642, Japan}}

\newcommand{\NRAO}{\affiliation{National Radio Astronomy Observatory, 520 Edgemont Road, Charlottesville, VA 22903-2475, USA}}

\newcommand{\OAN}{\affiliation{Observatorio Astron\'{o}mico Nacional (IGN), C/Alfonso XII, 3, E-28014 Madrid, Spain}}

\newcommand{\ObsParis}{\affiliation{Sorbonne Universit\'{e}, Observatoire de Paris, Universit\'{e} PSL, CNRS, LERMA, F-75014, Paris, France}}

\newcommand{\Oxford}{\affiliation{Sub-department of Astrophysics, Department of Physics, University of Oxford, Keble Road, Oxford OX1 3RH, UK}}

\newcommand{\Princeton}{\affiliation{Department of Astrophysical Sciences, Princeton University, Princeton, NJ 08544 USA}}

\newcommand{\STScI}{\affiliation{Space Telescope Science Institute, 3700 San Martin Drive, Baltimore, MD 21218, USA}}

\newcommand{\Sydney}{\affiliation{Sydney Institute for Astronomy, School of Physics A28, The University of Sydney, NSW 2006, Australia}}

\newcommand{\TAPIR}{\affil{TAPIR, California Institute of Technology, Pasadena, CA 91125, USA}}

\newcommand{\Tamkang}{\affiliation{Department of Physics, Tamkang University, No.151, Yingzhuan Rd., Tamsui Dist., New Taipei City 251301, Taiwan}}

\newcommand{\Toulouse}{\affiliation{Universit\'{e} de Toulouse, UPS-OMP, IRAP, F-31028 Toulouse cedex 4, France}}

\newcommand{\Toledo}{\affiliation{University of Toledo, 2801 W. Bancroft St., Mail Stop 111, Toledo, OH 43606, USA}}

\newcommand{\UChile}{\affiliation{Departamento de Astronom\'{i}a, Universidad de Chile, Camino del Observatorio 1515, Las Condes, Santiago, Chile}}

\newcommand{\UCM}{\affiliation{Departamento de F\'{\i}sica de la Tierra y Astrof\'{\i}sica, Universidad Complutense de Madrid, E-28040, Spain}}

\newcommand{\UCSD}{\affiliation{Center for Astrophysics and Space Sciences, Department of Physics,  University of California, San Diego, 9500 Gilman Drive, La Jolla, CA 92093, USA}}

\newcommand{\UMass}{\affiliation{University of Massachusetts—Amherst, 710 N. Pleasant Street, Amherst, MA 01003, USA}}

\newcommand{\Wyoming}{\affiliation{Department of Physics and Astronomy, University of Wyoming, Laramie, WY 82071, USA}}

\newcommand{\Zurich}{\affiliation{Institute for Computational Science, University of Z\"urich, Winterthurerstrasse 190, 8057 Z\"urich, Switzerland}}


\author[0000-0003-0378-4667]{Jiayi~Sun \begin{CJK*}{UTF8}{gbsn}(孙嘉懿)\end{CJK*}}
\altaffiliation{CITA National Fellow}
\McMaster
\CITA

\author[0000-0002-2545-1700]{Adam~K.~Leroy}
\OSU
\CCAPP

\author[0000-0002-0509-9113]{Eve~C.~Ostriker}
\Princeton

\author[0000-0002-6118-4048]{Sharon~Meidt}
\Gent

\author[0000-0002-5204-2259]{Erik~Rosolowsky}
\Alberta

\author[0000-0002-3933-7677]{Eva~Schinnerer}
\MPIA

\author[0000-0001-5817-0991]{Christine~D.~Wilson}
\McMaster

\author[0000-0003-4161-2639]{Dyas~Utomo}
\NRAO

\author[0000-0002-2545-5752]{Francesco~Belfiore}
\INAF

\author[0000-0003-4218-3944]{Guillermo~A.~Blanc}
\Carnegie
\UChile

\author[0000-0002-6155-7166]{Eric~Emsellem}
\ESO
\Lyon

\author[0000-0001-5310-467X]{Christopher~Faesi}
\UMass

\author[0000-0002-9768-0246]{Brent~Groves}
\ICRAR

\author[0000-0002-9181-1161]{Annie~Hughes}
\CNRS

\author[0000-0001-9605-780X]{Eric~W.~Koch}
\CfA

\author[0000-0001-6551-3091]{Kathryn~Kreckel}
\Heidelberg

\author[0000-0001-9773-7479]{Daizhong~Liu}
\MPE

\author[0000-0002-1370-6964]{Hsi-An~Pan}
\Tamkang

\author[0000-0003-3061-6546]{J\'er\^ome~Pety}
\IRAM
\ObsParis

\author[0000-0002-0472-1011]{Miguel~Querejeta}
\OAN

\author[0000-0001-7876-1713]{Alessandro~Razza}
\UChile

\author[0000-0002-2501-9328]{Toshiki~Saito}
\NAOJ

\author[0000-0002-5783-145X]{Amy~Sardone}
\OSU 
\CCAPP

\author[0000-0003-1242-505X]{Antonio~Usero}
\OAN

\author[0000-0002-0012-2142]{Thomas~G.~Williams}
\Oxford
\MPIA

\author[0000-0003-0166-9745]{Frank~Bigiel}
\Bonn

\author[0000-0002-5480-5686]{Alberto~D.~Bolatto}
\Maryland

\author[0000-0002-5635-5180]{M\'elanie~Chevance}
\ITA
\COOL

\author[0000-0002-5782-9093]{Daniel~A.~Dale}
\Wyoming

\author[0000-0001-6119-9883]{Jindra~Gensior}
\Zurich

\author[0000-0001-6708-1317]{Simon~C.~O.~Glover}
\ITA

\author[0000-0002-3247-5321]{Kathryn~Grasha}
\ANU
\ASTROThreeD

\author[0000-0001-9656-7682]{Jonathan~D.~Henshaw}
\Liverpool
\MPIA

\author[0000-0002-9165-8080]{Mar\'ia~J.~Jim\'{e}nez-Donaire}
\OAN

\author[0000-0002-0560-3172]{Ralf~S.~Klessen}
\ITA
\IWR

\author[0000-0002-8804-0212]{J.~M.~Diederik~Kruijssen}
\COOL

\author[0000-0001-7089-7325]{Eric~J.~Murphy}
\NRAO

\author[0000-0001-9793-6400]{Lukas~Neumann}
\Bonn

\author[0000-0003-4209-1599]{Yu-Hsuan~Teng}
\UCSD

\author[0000-0002-8528-7340]{David~A.~Thilker}
\JHU

\correspondingauthor{Jiayi~Sun}
\email{sun208@mcmaster.ca}


\begin{abstract}
We measure empirical relationships between the local star formation rate (SFR) and properties of the star-forming molecular gas on 1.5~kpc scales across 80 nearby galaxies.
These relationships, commonly referred to as ``star formation laws,'' aim at predicting the local SFR surface density from various combinations of molecular gas surface density, galactic orbital time, molecular cloud free-fall time, and the interstellar medium dynamical equilibrium pressure.
Leveraging a multiwavelength database built for the PHANGS survey, we measure these quantities consistently across all galaxies and quantify systematic uncertainties stemming from choices of SFR calibrations and the CO-to-H$_2$ conversion factors.
The star formation laws we examine show 0.3--0.4~dex of intrinsic scatter, among which the molecular Kennicutt-Schmidt relation shows a $\sim$10\% larger scatter than the other three.
The slope of this relation ranges $\beta\approx0.9{-}1.2$, implying that the molecular gas depletion time remains roughly constant across the environments probed in our sample.
The other relations have shallower slopes ($\beta\approx0.6{-}1.0$), suggesting that the star formation efficiency (SFE) per orbital time, the SFE per free-fall time, and the pressure-to-SFR surface density ratio (i.e., the feedback yield) vary systematically with local molecular gas and SFR surface densities.
Last but not least, the shapes of the star formation laws depend sensitively on methodological choices.
Different choices of SFR calibrations can introduce systematic uncertainties of at least 10--15\% in the star formation law slopes and 0.15--0.25~dex in their normalization, while the CO-to-H$_2$ conversion factors can additionally produce uncertainties of 20--25\% for the slope and 0.10--0.20~dex for the normalization.
\end{abstract}


\section{Introduction} \label{sec:intro}

``Star formation laws'' are empirical scaling relations between properties of the interstellar gas and the star formation rate (SFR) of this gas.
These relations arise from the physical processes governing star formation in the interstellar medium (ISM) in galaxies near and far \citep[see review by][]{Kennicutt_Evans_2012}.

Since the pioneering work of \citet{Schmidt_1959}, many forms of star formation laws (``SF laws'' hereafter) have been proposed in the literature. 
One large family is known as ``integrated'' SF laws, which connect unresolved, global measurements of galaxy gas mass and SFR \citep[e.g.,][]{Kennicutt_1998a,Saintonge_etal_2011a,delosReyes_Kennicutt_2019}.
The other family describes ``resolved'' SF laws, which relate the local surface densities of gas mass and SFR \citep[usually measured at $\sim$kpc scales; e.g.,][]{Wong_Blitz_2002,Bigiel_etal_2008}.
Alternative formulations have modified the basic relationship by, e.g.,  (a) including only molecular gas \citep[e.g.,][]{Wong_Blitz_2002,Bigiel_etal_2011} or dense molecular gas \citep[e.g.,][]{Gao_Solomon_2004,Lada_etal_2012} as opposed to the total neutral gas, (b) considering volume densities instead of surface densities \citep[e.g.,][]{Schmidt_1959,Bacchini_etal_2019a}, or (c) incorporating additional information beyond gas mass/densities for the independent variable \citep[such as orbital time, see][]{Elmegreen_1997,Silk_1997}.
The functional forms of some of the resolved empirical SF laws are originally motivated by theoretical considerations, as we shall discuss below.

Among the resolved SF laws, at least four of them have attracted greater attention in recent decades.
These are:
(1) the molecular Kennicutt--Schmidt relation \citep{Kennicutt_1998a} between the surface densities of molecular gas and SFR;
(2) the molecular Elmegreen--Silk relation \citep{Elmegreen_1997,Silk_1997} between the SFR surface density and molecular gas surface density divided by orbital time;
(3) the free-fall time regulated SF relation \citep{McKee_Ostriker_2007,Krumholz_etal_2009a,Krumholz_etal_2012} linking the SFR surface density to the molecular gas surface density divided by the molecular cloud free-fall time;
and (4) the pressure-regulated SF relation \citep{Ostriker_etal_2010,Ostriker_Shetty_2011} connecting the SFR surface density to the ISM dynamical equilibrium pressure.
Many works have suggested near-unity slopes for these relations \citep[between 0.8 and 1.2; see e.g.,][]{Daddi_etal_2010,Genzel_etal_2010,Bigiel_etal_2011,Krumholz_etal_2012,Ostriker_Kim_2022}, which implies that the ratio of the dependent and independent variables (i.e., the proportionality constant) remains roughly unchanged across a wide range of physical conditions.
These SF laws and the corresponding proportionality constants (namely the molecular gas depletion time, the star formation efficiency per unit orbital time and per unit free-fall time, and the feedback yield) are thus of great interest and have been linked to various star formation theories.

In this Letter, we intend to accomplish three overarching goals.
First, we aim to provide the latest measurements of these four SF laws and their associated proportionality constants across 80 nearby, star-forming galaxies mapped by the PHANGS--ALMA survey \citep{Leroy_etal_2021a,Leroy_etal_2021b}.
The excellent depth, resolution, and field-of-view coverage of the PHANGS--ALMA survey allow us to measure these relations throughout a representative range of star-forming environments in the local universe.
Second, we compare these star formation laws in terms of their overall slopes and scatter, as well as the scatter in the corresponding proportionality constants.
These measurements provide an empirical basis for determining the predictive power of these SF laws for the local SFR.
Third, we examine how the best-fit slope, normalization, and scatter of each SF law depends on the approach used to estimate physical quantities, especially the SFR and molecular gas mass, from observable quantities.
Such estimation often relies on various methodological choices (such as a particular SFR calibration or CO-to-H$_2$ conversion factor), and many different choices have been proposed for different physical regimes or under different observational limitations.
Quantitative comparisons among even a subset of these choices can provide useful estimates for the systematic uncertainties they introduce.

We note that the molecular Kennicutt--Schmidt relation has been reported for PHANGS galaxies in various sub-samples and sub-galactic environments \citep[see][]{Pessa_etal_2021,Querejeta_etal_2021}; the star formation efficiency per free-fall time has been measured within the PHANGS--ALMA pilot sample \citep{Utomo_etal_2018}; and the pressure-regulated SF relation has been presented for a subset of PHANGS targets \citep{Sun_etal_2020b}.
This Letter provides updated measurements across the full PHANGS--ALMA sample, leveraging the latest processing of the same observational datasets used in \citet{Utomo_etal_2018}, \citet{Sun_etal_2020b}, and \citet{Querejeta_etal_2021}, as well as applying refined methodologies for converting observable quantities into physical quantities.


\section{Data} \label{sec:data}

\defcitealias{Sun_etal_2022}{S22}

\defcitealias{Bolatto_etal_2013}{B13}
\defcitealias{Sun_etal_2020b}{S20}
\defcitealias{Gong_etal_2020}{G20}

We base our analysis on the PHANGS high-level measurement database described in \citet[hereafter \citetalias{Sun_etal_2022}]{Sun_etal_2022}.
The database incorporates multiwavelength data for 80 galaxies, extracts observational measurements and associated uncertainties with matched sampling and weighting schemes, and converts them into physical quantities following a set of best practices.
In this paper, we use the latest version of this database, which sees many improvements over the version published in \citetalias{Sun_etal_2022}.
We summarize these improvements and announce the online release of the associated data products in Appendix~\ref{apdx:mega-table}.

Below we list the key physical quantities used in this paper and describe their data sources as well as the methodological choices and assumptions involved in their derivation.

\begin{itemize}[itemsep=0.5em,leftmargin=1em,parsep=0em,partopsep=0em]

\item[$\sbullet$] \emph{Star formation rate surface density, $\SigSFR$}.
We derive this quantity at a fixed 1.5~kpc resolution\footnote{This is the best common resolution achievable for all galaxies in our sample since we rely on WISE 22~$\mu$m data.} from three different SFR calibrations that combine UV, optical, and/or IR data (see Table~\ref{tab:setup}).
Our fiducial choice is to combine narrow-band H$\alpha$ data (acquired with the 2.5m du~Pont Telescope and the ESO/MPG 2.2m Telescope; A.~Razza et al.\ in preparation) with WISE 22~$\mu$m data \citep{Leroy_etal_2019}.
For this combination, we use a new SFR calibration proposed by \citet{Belfiore_etal_2023}, which can better mitigate contamination from IR cirrus in the 22~$\mu$m band than the classic \citet{Calzetti_etal_2007} calibration.
Alternatively, we combine GALEX far-UV data \citep[154~nm;][]{Leroy_etal_2019} with WISE 22~$\mu$m data, following another new calibration proposed by \citet{Belfiore_etal_2023} for similar purposes.
Finally, for 19 galaxies in the PHANGS-MUSE sample \citep{Emsellem_etal_2022}, we also include H$\alpha$-based SFR measurements that are corrected for dust extinction based on the Balmer decrement \citep[also see \citealt{Pessa_etal_2021}]{Belfiore_etal_2023}.
This last calibration is likely the most reliable (because of the trustworthy [\ion{N}{2}] subtraction, extinction correction, and the superior depth of the MUSE observations), but unfortunately, the required MUSE data are available for only $\sim$1/4 of the galaxies studied here.
We note that our H$\alpha$-based SFR calculations include all diffuse ionized gas emission, as recommended for kpc-scale observations of star-forming galaxies \citep{Belfiore_etal_2022a}.
All our $\SigSFR$ measurements have been corrected for galaxy inclination (as described in \citetalias{Sun_etal_2022}).

\item[$\sbullet$] \emph{Molecular gas surface density, $\Sigmol$}.
We derive this quantity from PHANGS-ALMA \CO21\ data \citep{Leroy_etal_2021a,Leroy_etal_2021b} at 1.5~kpc resolution.
More specifically, we use \CO21\ integrated intensities from the high-completeness, ``broad'' moment-0 maps \citep[see][for more details]{Leroy_etal_2021b} and convert them into molecular gas surface densities using four different prescriptions for the CO-to-H$_2$ conversion factor ($\alphaCO$, also see Table~\ref{tab:setup}).
Our fiducial choice is to combine a varying, metallicity-dependent $\alpha_\mathrm{CO\,(1{-}0)}$ with a fixed CO line ratio of $R_{21}=0.65$ \citep[but see Section~\ref{sec:results:methods} for some caveats]{Leroy_etal_2022}. 
Here, the metallicity is inferred from empirical scaling relations and ranges 0.6--1.2 solar for most regions studied in this work
\citep[see][hereafter \citetalias{Sun_etal_2020b}]{Sun_etal_2020b}.
We also employ three alternative $\alphaCO$ treatments: (1) a fixed, Galactic $\alpha_\mathrm{CO\,(1{-}0)}$ value of $4.35\;\ualphaCO$; (2) an empirical $\alpha_\mathrm{CO\,(1{-}0)}$ calibration depending on metallicity and total (gas + star) mass surface density\footnote{While the original \citetalias{Bolatto_etal_2013} prescription also includes a molecular cloud surface density term and it is implemented as such in \citetalias{Sun_etal_2020b} and \citetalias{Sun_etal_2022}, we have noticed that it often leads to unphysical $\alphaCO$ values in low surface density regimes. Here we use a fixed cloud surface density of $100\,\uSig$ to mitigate this issue.} \citep[hereafter \citetalias{Bolatto_etal_2013}]{Bolatto_etal_2013}; and (3) a simulation-motivated $\alpha_\mathrm{CO\,(2{-}1)}$ prescription relying on metallicity, CO integrated intensity, and resolution of the CO measurements\footnote{We first calculate the \citetalias{Gong_etal_2020} $\alphaCO$ based on 150~pc resolution CO data, then spatially average it to 1.5~kpc resolution \citep[see Appendix~B in][]{Sun_etal_2022}.} \citep[hereafter \citetalias{Gong_etal_2020}]{Gong_etal_2020}.
The last two options are likely the most realistic given the additional physics they intend to capture (e.g., gas excitation, see Section~\ref{sec:results:methods}).
Details on the implementation of these $\alphaCO$ prescriptions can be found in \citetalias{Sun_etal_2022}.
The $\Sigmol$ values are also corrected for galaxy inclination.

\begin{deluxetable}{lll}
\tablecaption{Methodological Choices\label{tab:setup}}
\tablewidth{0pt}
\tablehead{
\colhead{Method shorthand} &
\colhead{$\alphaCO$} &
\colhead{SFR calibration} \\[-1.5em]
}
\startdata
\\[-1.5em]
``Fiducial'' & \citetalias{Sun_etal_2020b} & H$\alpha$+22\micron \\
``FUV+W4 SFR'' & \citetalias{Sun_etal_2020b} & FUV+22\micron \\
``Av-corr H$\alpha$ SFR'' & \citetalias{Sun_etal_2020b} & $A_V$-corrected H$\alpha$\tablenotemark{$\star$} \\
``MW $\alpha_\mathrm{CO}$'' & Galactic & H$\alpha$+22\micron \\
``B13 $\alpha_\mathrm{CO}$'' & \citetalias{Bolatto_etal_2013} & H$\alpha$+22\micron \\
``G20 $\alpha_\mathrm{CO}$'' & \citetalias{Gong_etal_2020} & H$\alpha$+22\micron \\
\enddata
\tablenotetext{\star}{Available only for 19 galaxies in the PHANGS-MUSE sample \citep{Emsellem_etal_2022}.}
\tablecomments{
Reference for all SFR calibrations used here: \citet{Belfiore_etal_2023}.
}
\vspace{-2\baselineskip}
\end{deluxetable}
\vspace{-1\baselineskip}

\item[$\sbullet$] \emph{Galactic orbital time, $\torb$}.
We derive this quantity from the rotation curve models based on \CO21\ kinematics presented in \citep{Lang_etal_2020}, which exist for 62 out of our 80 galaxies.
As discussed in \citetalias{Sun_etal_2022}, we use a set of parametrized model fits to the measured rotation curves (J.~Nofech et al.\ in preparation), which effectively suppress the fluctuation of rotational velocities across radial bins (due to, e.g., non-circular motions).
The orbital time is then determined from the galactocentric radius and the local circular velocity given by the rotation curve models.

\item[$\sbullet$] \emph{Population-averaged molecular cloud free-fall time, $\tff$}.
We derive this quantity for each 1.5~kpc region by calculating the mass-weighted harmonic mean of the free-fall time of all molecular clouds located in that region, as described in \citetalias{Sun_etal_2022}.
In this work, we use the molecular cloud free-fall time measured from the 150~pc scale \CO21\ maps, denoted as $\tffsub{150pc}$.
This measurement is available for all 80 galaxies.
It adopts a simplifying assumption that the emission in each 150~pc beam originates from a beam-filling, spherical cloud \citep[consistent with][]{Sun_etal_2018,Sun_etal_2020a}.
We adopt the same conversion factor for $\tff$ as for $\Sigmol$.

\item[$\sbullet$] \emph{ISM dynamical equilibrium pressure, $P\sbsc{DE}$}.
We derive this quantity on 1.5~kpc scale by combining the total gas surface density $\Siggas=\Sigmol+\Sigatom$, stellar mass volume density at the disk mid-plane $\rho_\star$, and vertical gas velocity dispersion $\sigma_\mathrm{gas,\,z}$, closely following \citetalias{Sun_etal_2020b} (also see \citealt{Ostriker_etal_2010,Ostriker_Kim_2022}):
\vspace{-0.3\baselineskip}
\begin{equation}
P\sbsc{DE} =
\frac{\pi G}{2}\,\Siggas^2 + \Siggas\,\sqrt{2G\rho_\star}\,\sigma_\mathrm{gas,\,z}~.
\label{eq:P_DE}
\end{equation}
Here we calculate $\Sigatom$ from \HI\ 21~cm line data gathered by various observing programs on VLA, ATCA, and WSRT (see \citetalias{Sun_etal_2022} for a full list).
We estimate $\rho_\star$ by first calculating the 2D stellar mass surface density $\Sigma_\star$ from near-IR data gathered by WISE and Spitzer (with a locally determined mass-to-light ratio; see \citealt{Leroy_etal_2021a}), and then converting it to stellar volume density assuming the stellar disk thickness scales with its radial extent \citep[see][\citetalias{Sun_etal_2020b}]{Kregel_etal_2002}.
We adopt a fixed $\sigma_\mathrm{gas,\,z}=11\;\uV$ following \citet[but see \citetalias{Sun_etal_2020b} and \citealt{Ostriker_Kim_2022} for discussions about the caveats related to this assumption]{Ostriker_etal_2010}.
In total, we are able to measure $\PDE$ in 48 out of 80 galaxies, with the sample size limited primarily by the availability of \HI\ data.

\end{itemize}


\section{Results} \label{sec:results}

With all the key physical quantities listed in Section~\ref{sec:data}, we examine the four star formation laws described in Section~\ref{sec:intro} across the full PHANGS--ALMA sample.
With over 2,000 kpc-size regions across 80 galaxies, this is the largest sample for which all the necessary quantities (including the orbital time, cloud-scale free-fall time, and the ISM dynamical equilibrium pressure) can be measured directly from observations.
Our uniform methodological treatments across the full sample allow for rigorous comparisons between the four star formation laws, as well as systematic explorations of how particular methodological choices (Table~\ref{tab:setup}) influence the quantitative results.

\subsection{Molecular Kennicutt--Schmidt Relation} \label{sec:results:mKS}

\begin{figure*}
\gridline{\fig{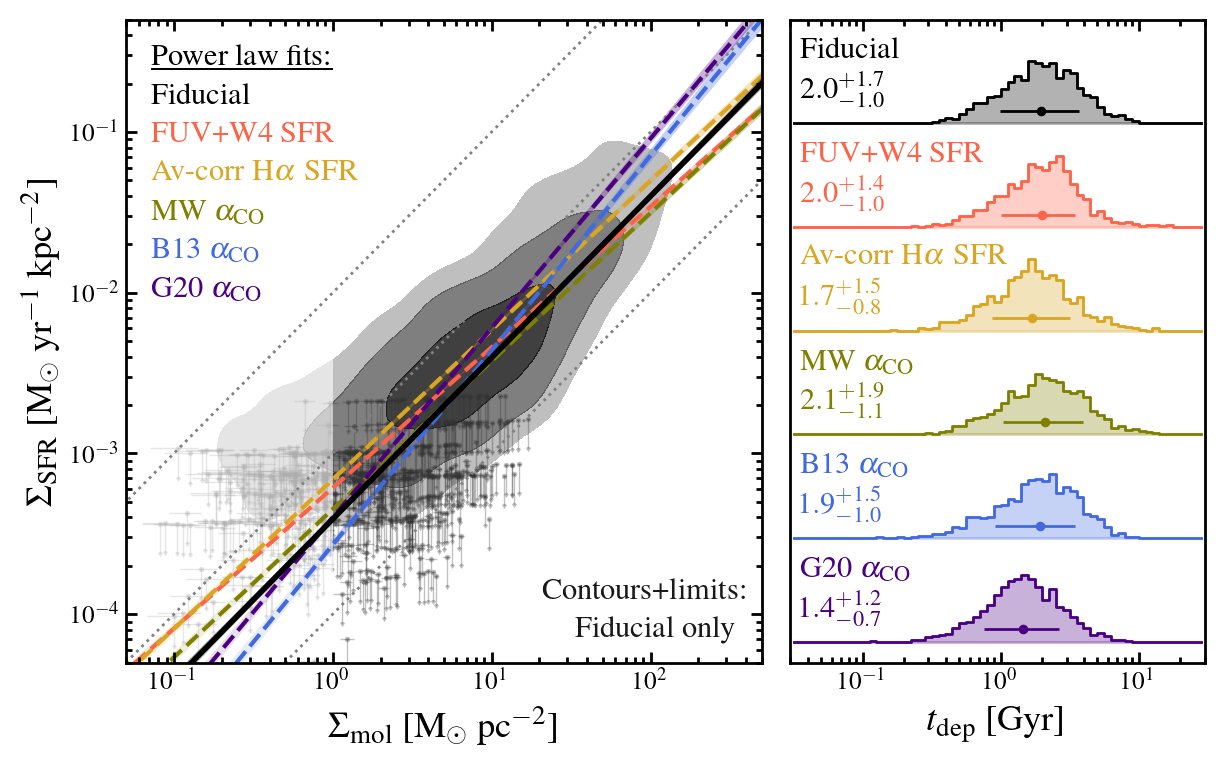}{0.85\textwidth}{}}
\vspace{-2.5\baselineskip}
\caption{
\textit{Left panel:} The molecular Kennicutt--Schmidt (mKS) relation across the PHANGS--ALMA sample.
The density contours (40\%--80\%--95\% levels) show the distributions of all 1.5~kpc scale regions with $>$3$\sigma$ detections for both $\Sigmol$ and $\SigSFR$, and the downward arrows show 3$\sigma$ upper limits for $\SigSFR$ (see Section~\ref{sec:results}).
The solid black line shows the best-fit power law model for all detections and upper limits above $\Sigmol=1\;\uSig$ (i.e., the non-shaded side for the contours and symbols), where there is minimal censoring on $\Sigmol$.
The other colored lines show the best-fit model when using alternative SFR calibrations or CO-to-H$_2$ conversion factors.
The thin dotted lines mark linear relations with constant molecular gas depletion times of 0.1, 1, and 10~Gyr (top-left to bottom-right).
\textit{Right panel:}
Normalized histograms of the molecular gas depletion time $\tdep=\Sigmol/\SigSFR$, color-coded similarly to the left panel.
The median value and 16$^{\rm th}$--84$^{\rm th}$ percentile range are marked by a dot and a horizontal bar beneath each histogram, with their values displayed to the left of each histogram.
}
\vspace{0.5\baselineskip}
\label{fig:mKS}
\end{figure*}

We first examine the relationship between surface densities of molecular gas mass and SFR, commonly known as the molecular Kennicutt--Schmidt (mKS) relation.
While the original KS relation uses the total gas surface density as the independent variable \citep{Kennicutt_1989}, it has been shown that the relation with $\Sigmol$ is tighter and has a more consistent slope across diverse environments \citep[e.g.,][]{Wong_Blitz_2002,Bigiel_etal_2011}.
This slope is often found to be close to unity in local star-forming galaxies.
As a result, the molecular gas depletion time $\tdep \equiv \Sigmol / \SigSFR$ varies only weakly, with typical values of $1\text{--}3$~Gyr \citep[e.g.,][also see review by \citealt{Saintonge_Catinella_2022}]{Leroy_etal_2008,Saintonge_etal_2011a}.

Figure~\ref{fig:mKS} shows the mKS relation measured at 1.5~kpc scale (left panel) and the corresponding distribution of $\tdep$ (right panel) across the full PHANGS--ALMA sample.
With the fiducial methodological choices (see Table~\ref{tab:setup}), our measurements span three decades in $\Sigmol$ ($10^{-1}\text{--}10^2\,\uSig$) and $\SigSFR$ ($10^{-4}\text{--}10^{-1}\,\uSigSFR$).
The corresponding $\tdep$ distribution shows a median value and a $\pm1\sigma$ range of $2.0^{+1.7}_{-1.0}$~Gyr.
A large fraction of our measurements are 3$\sigma$ upper limits\footnote{Upper limits of $\Sigmol$ are omitted in all figures for clarity.} at $\Sigmol<1\,\uSig$ or $\SigSFR<10^{-3}\,\uSigSFR$ due to the finite sensitivities of the CO, H$\alpha$, or IR observations.
The distributions of detections and 3$\sigma$ upper limits in the $\Sigmol$--$\SigSFR$ space also vary moderately depending on the choice of SFR calibrations and $\alphaCO$ prescriptions.

To further quantify the shape and tightness of the mKS relation, we fit a power law model to the data distribution with a functional form of
\begin{align}
\log_{10}&\!\left(\frac{\SigSFR}{\uSigSFR}\right)\nonumber\\
&= \alpha+\beta\log_{10}\!\left(\frac{\Sigmol}{10\;\uSig}\right)~,
\label{eq:mKS}
\end{align}
where the normalization $\alpha$ is determined at $\Sigmol=10\;\uSig$ (close to the mid-point of our sample).
The model fit is performed in logarithmic space with the \texttt{linmix} package \citep{Kelly_2007}.
It determines the power law normalization ($\alpha$), slope ($\beta$), and the intrinsic scatter ($\sigma$) around it from the data distribution, accounting for measurement uncertainties and non-detections for the dependent variable.
We further restrict the fit to measurements above a $\Sigmol$ threshold in order to minimize biases caused by non-detections for the independent variable.
This threshold is $\Sigmol=1\,\uSig$ for the fiducial $\alphaCO$ and varies between $1$--$2\,\uSig$ for different $\alphaCO$ choices (see Appendix~\ref{apdx:fit}).

The first part of Table~\ref{tab:fit} reports the best-fit model parameters for the mKS relation when adopting various SFR calibrations and $\alphaCO$.
In all cases, we see near-unity power law indices ($\beta=0.88$--$1.21$) and small intrinsic scatters ($\sigma=0.29$--$0.38$~dex).
The near-unity slope means the scatter in $\tdep$ (about a factor of two) is almost identical to the residual scatter around the best-fit power law relation.
Considering the impact of SFR calibration and $\alphaCO$ choices separately, we find the former can change the slope by 13\% and the normalization by 0.17~dex, whereas the latter produces changes of 25\% for the slope and 0.18~dex for the normalization.
These findings call special attention to the methodology-dependent nature of the mKS relation shape and slope.

\subsection{Molecular Elmegreen--Silk Relation} \label{sec:results:mES}

\begin{figure*}
\gridline{\fig{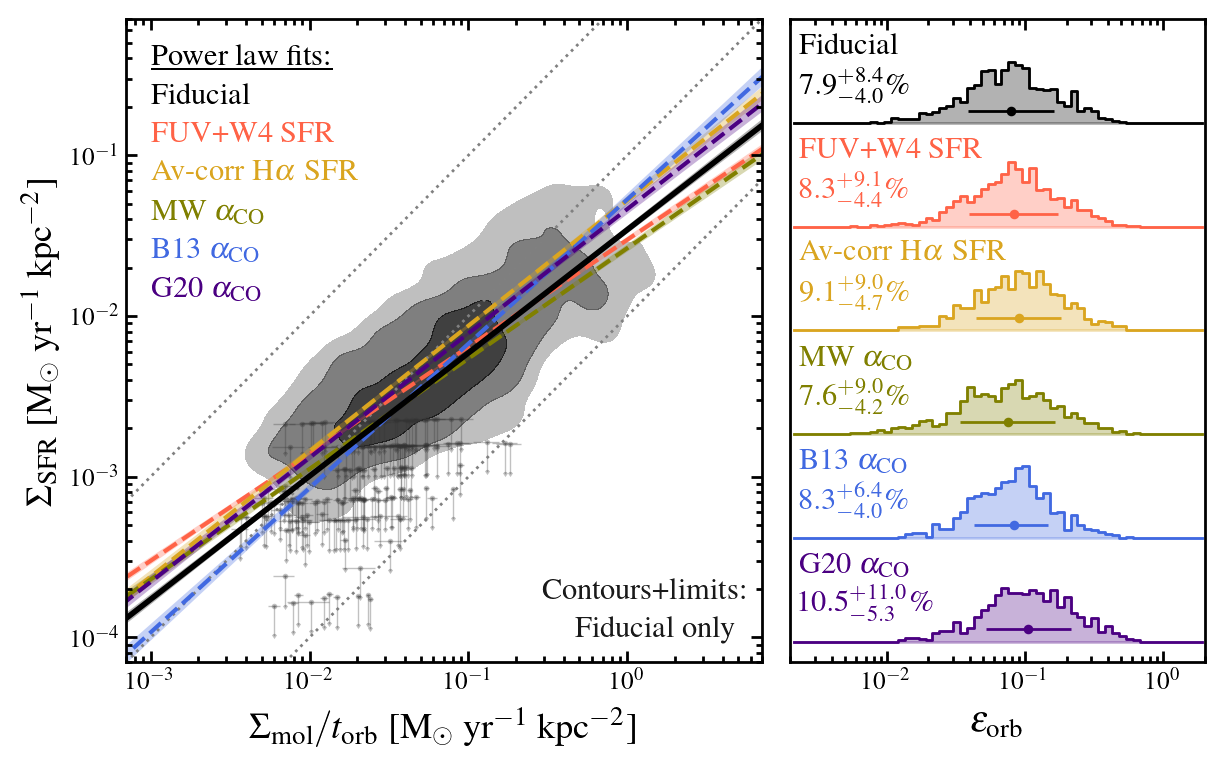}{0.85\textwidth}{}}
\vspace{-2.5\baselineskip}
\caption{
Similar to Figure~\ref{fig:mKS}, but here showing the molecular Elmegreen--Silk relation (mES relation; \textit{left panel}) and normalized histograms of the star formation efficiency per orbital time, $\SFEorb=(\SigSFR/\Sigmol)\,\torb$ (\textit{right panel}).
The thin dotted lines in the left panel mark linear relations with constant $\SFEorb$ of 1\%, 10\%, and 100\% (bottom-right to top-left).
Note that data below the $\Sigmol$ threshold shown in Figure~\ref{fig:mKS} are also excluded in the power law fit for the mES (and all other) relations.
}
\label{fig:mES}
\end{figure*}

The KS relation links the current SFR to the amount of gas available at the moment.
If there is a characteristic timescale on which the gas is converted into stars, then a potentially more direct (and more physical) relation would connect the current SFR to the amount of gas \textit{normalized by that timescale}.
The molecular Elmegreen--Silk relation \citep[mES relation,][]{Elmegreen_1997,Silk_1997} is one such candidate, for which the normalizing timescale is the galactic orbital time, $\torb$.
This timescale is relevant to not only galactic rotation, but also shear, spiral arm passages, and cloud--cloud collisions, all of which can regulate cloud formation/destruction and in this way denote the rlevant timescale for star formation \citep[e.g.,][]{Tan_2000}.
In this case, the independent variable becomes $\Sigmol/\torb$, and the ratio between $\SigSFR$ and this new independent variable defines the star formation efficiency per orbital time, $\SFEorb=(\SigSFR/\Sigmol)\,\torb$.

Figure~\ref{fig:mES} shows the mES relation (left panel) and the distribution of $\SFEorb$ (right panel).
These measurements are available for a subsample of 62 galaxies, for which we can determine $\torb$ from rotation curve models (see Section~\ref{sec:data}).
In addition, it is often challenging to measure the rotation curve near the edge of the CO images, due to incomplete azimuthal coverage and sparse CO detection.
Consequently, there are visibly fewer measurements at the low $\SigSFR$ end in Figure~\ref{fig:mES} than in Figure~\ref{fig:mKS}.

Within the range of environments where we do have an adequate number of measurements (i.e., $\SigSFR\sim10^{-3.5}$--$10^{-1}\,\uSigSFR$), we fit a power law model for the mES relation:
\begin{align}
\log_{10}&\!\left(\frac{\SigSFR}{\uSigSFR}\right)\nonumber\\
&= \alpha+\beta\log_{10}\!\left(\frac{\Sigmol/\torb}{0.1\;\uSigSFR}\right)~.
\label{eq:mES}
\end{align}
The mES relation has a much shallower slope than the mKS relation for any given methodological choice (Table~\ref{tab:fit}).
This is expected, as regions with higher $\Sigmol$ are typically located at smaller galactocentric radii and thus have shorter $\torb$.
In other words, the dynamic range in $\Sigmol/\torb$ is usually wider than that in $\Sigmol$ for the same set of regions, resulting in a shallower mES relation than the mKS relation.
While this is not obvious from Figure~\ref{fig:mKS} and \ref{fig:mES} due to the different samples of measurements they include, we have verified it in a common subsample (i.e., the intersection of the samples in Figure~\ref{fig:mKS} and \ref{fig:mES}).

The intrinsic scatter of the mES relation is also smaller than that of the mKS relation for any given methodological choice.
This suggests that the mES relation makes empirically better predictions for $\SigSFR$ than the mKS relation.
However, the sub-linear slope of the mES relation indicates that $\SFEorb$ decreases systematically towards the high $\SigSFR$ end and shows a wider distribution than $\tdep$ across the whole sample.
It is thus not a good assumption to adopt a fixed $\SFEorb=5$--$10\%$ across all regions, even though it could be reasonable to assume a fixed $\tdep$ for the same range of conditions.

\subsection{Free-fall Time Regulated SF Relation} \label{sec:results:FFTR}

\begin{figure*}
\gridline{\fig{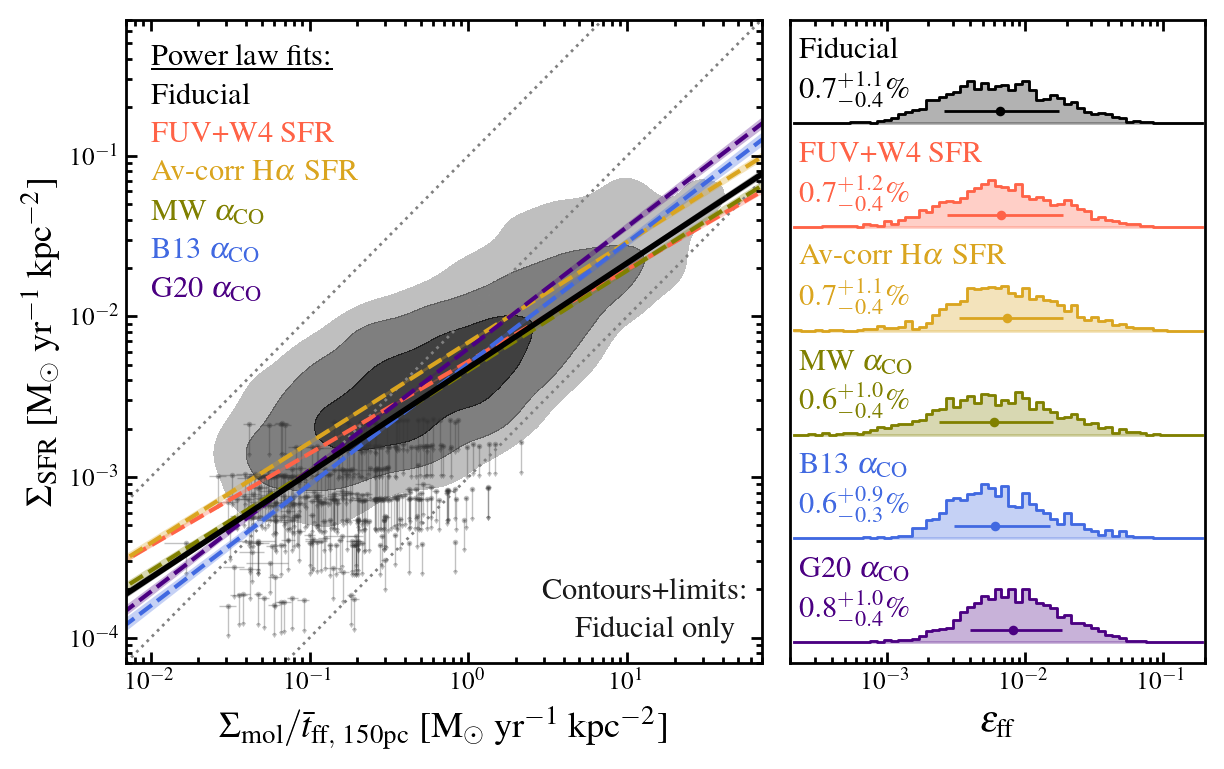}{0.85\textwidth}{}}
\vspace{-2.5\baselineskip}
\caption{
Similar to Figure~\ref{fig:mKS}, but here showing the free-fall time regulated SF relation (FFTR relation; \textit{left panel}) and normalized histograms of the star formation efficiency per free-fall time, $\SFEff=(\SigSFR/\Sigmol)\,\tff$ (\textit{right panel}).
The thin dotted lines in the left panel mark linear relations with constant $\SFEff$ of 0.1\%, 1\%, and 10\% (bottom-right to top-left).
\vspace{-0.5\baselineskip}
}
\label{fig:FFTR}
\end{figure*}

Aside from the galactic orbital time, another highly relevant timescale for star formation is the free-fall time of molecular clouds, especially since molecular clouds are the immediate sites of star formation.
The free-fall time regulated SF relation  (FFTR relation) builds on this notion and connects the local SFR to the ratio of $\Sigmol$ and the average free-fall time of molecular clouds, $\tff$, in the same region \citep[e.g.,][]{Krumholz_etal_2009a,Krumholz_etal_2012}.
The star formation efficiency per free-fall time, $\SFEff=(\SigSFR/\Sigmol)\,\tff$, then describes the fraction of gas mass converted to stars over a unity $\tff$.
This parameter is of particular interest to both observers and theorists, as it can be determined from observable quantities \citep[e.g.,][see \citealt{Krumholz_etal_2019} for a compilation]{Utomo_etal_2018,Evans_etal_2022} and predicted from analytical and numerical models of turbulence-regulated star formation at cloud scales \citep[e.g.,][]{Krumholz_McKee_2005,Hennebelle_Chabrier_2011,Federrath_Klessen_2012,Padoan_etal_2012,Padoan_etal_2014,KimJG_etal_2020}.

Figure~\ref{fig:FFTR} shows the FFTR relation (left panel) and the distribution of $\SFEff$ (right panel) across our full sample of 80 galaxies.
Our measurements span a similar range in $\SigSFR$ here as in Figure~\ref{fig:mKS}.
However, regions with low $\Sigmol$ often do not have $\tff$ measurements, as they require (a) detecting individual molecular clouds in CO and (b) having enough clouds in the 1.5~kpc region to determine a population average.
The range of environments we can probe for the FFTR relation ends up being similar to those used in the power law fit for the mKS relation (i.e., those above the $\Sigmol$ threshold; see Section~\ref{sec:results:mKS}).

For the FFTR relation, we fit a power law model as
\begin{align}
\log_{10}&\!\left(\frac{\SigSFR}{\uSigSFR}\right)\nonumber\\
&= \alpha+\beta\log_{10}\!\left(\frac{\Sigmol/\tff}{\uSigSFR}\right)~.
\label{eq:FFTR}
\end{align}
Similar to our findings for the mES relation, the slope of the FFTR relation is also consistently shallower than the mKS relation, regardless of methodological choices (Table~\ref{tab:fit}).
This reflects a systematic trend that the molecular cloud populations in high $\Sigmol$ regions have on average higher densities (e.g., \citetalias{Sun_etal_2022}), and consequently shorter free-fall times.
The slope we find for the FFTR relation is sub-linear in most cases, implying that $\SFEff$ drops substantially in higher surface density environments.

The FFTR relation exhibits an intrinsic scatter of $\sigma\approx0.3$~dex, which is comparable to the mES relation and mildly smaller than the mKS relation.
But the 1$\sigma$ dispersion of the $\SFEff$ distribution is wider, again due to the sub-linear FFTR relation slope.
That is, the measured range of $\SFEff=0.7^{+1.1}_{-0.4}\%$ (for the fiducial SFR calibration and $\alphaCO$) can be seen as the combined results of the FFTR relation intrinsic scatter \textit{plus} a systematic trend of decreasing $\SFEff$ with $\SigSFR$ (and $\Sigmol$).
For studies that rely on an assumed constant $\SFEff$ values to predict SFR (as is done in many galaxy simulations), it would be important to also account for this systematic trend.

In the context of turbulence-regulated SF models, variations in $\SFEff$ are considered to be driven by changes in the physical properties of individual star-forming molecular clouds \citep[e.g.,][]{Krumholz_McKee_2005,Federrath_Klessen_2012}.
Broadly speaking, $\SFEff$ would be higher for clouds with higher turbulent Mach number $\mathcal{M}$ (which is proportional to the turbulent velocity dispersion $\sigma\sbsc{turb}$) and lower virial parameter $\alpha\sbsc{vir}$ \citep[e.g., see Figure~1 in][]{Federrath_Klessen_2012}.
Since high $\Sigmol$ and $\SigSFR$ regions tend to host molecular clouds with larger $\sigma\sbsc{turb}$ and smaller $\alpha\sbsc{vir}$ (see Figure~5 in \citetalias{Sun_etal_2022}), we would then expect $\SFEff$ to be higher in those regions.
Yet this expectation appears inconsistent with the empirical trends found in this work and in previous studies \citep[e.g.,][but see \citealt{Barnes_etal_2017} for caveats]{Leroy_etal_2017a,Schruba_etal_2019}.
That being said, a more rigorous and thorough comparison with theoretical predictions is beyond the scope of this work and will be addressed in S.~Meidt et al.\ (in preparation).

\subsection{Pressure Regulated SF Relation} \label{sec:results:PR}

\begin{figure*}
\gridline{\fig{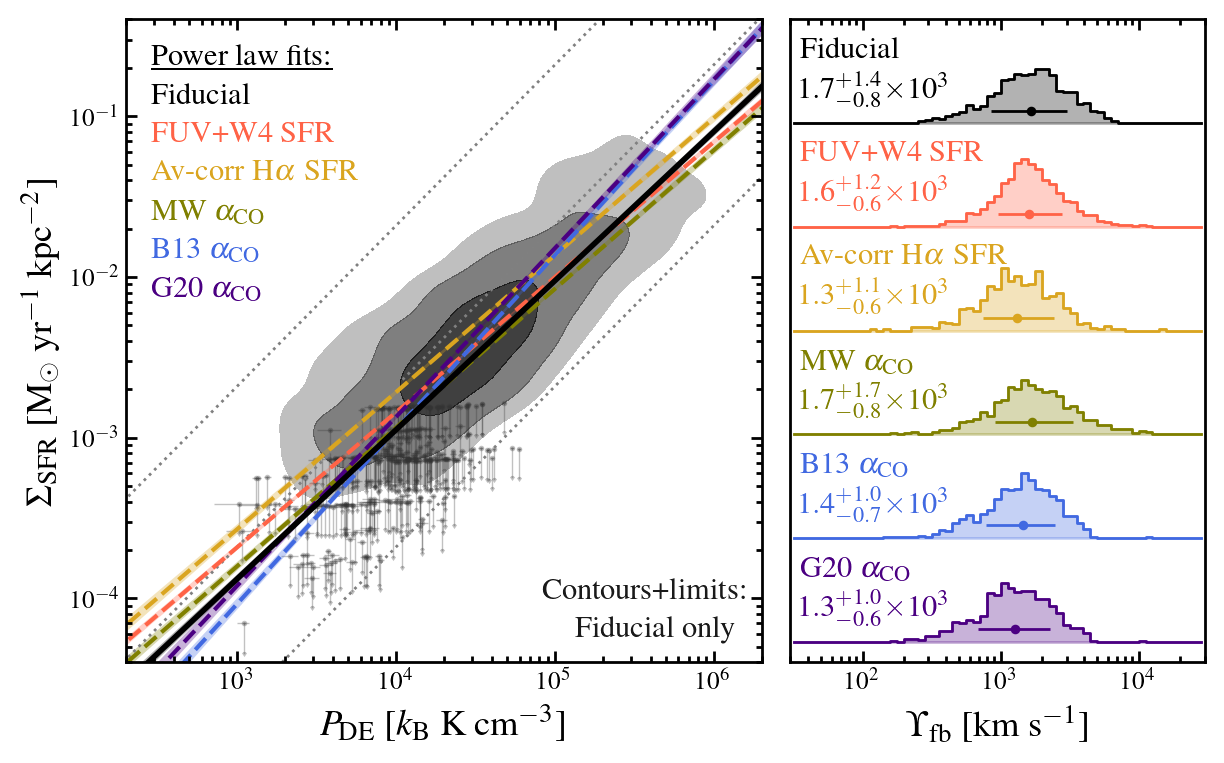}{0.85\textwidth}{}}
\vspace{-2.5\baselineskip}
\caption{
Similar to Figure~\ref{fig:mKS}, but here showing the pressure regulated SF relation (PR relation; \textit{left panel}) and normalized histograms of the feedback yield, $\Yfb=\PDE/\SigSFR$ (\textit{right panel}).
The thin dotted lines in the left panel mark linear relations with constant $\Yfb$ of $10^2$, $10^3$, and $10^4\;\uV$ (top-left to bottom-right).
}
\vspace{0.5\baselineskip}
\label{fig:PR}
\end{figure*}

The mES and FFTR relations discussed above measure the star formation efficiency relative to a specific dynamical timescale (either $\torb$ or $\tff$).
The focus, explicitly or implicitly, is on the ``mass supply'' aspect, with star formation thought of as a process that depletes the ISM.
The pressure-regulated, feedback-modulated star formation theory \citep{Ostriker_etal_2010,Ostriker_Kim_2022} instead views star formation as a source of energy and momentum, rather than a sink of mass, for the ISM.
In this framework, the local SFR determines the energy and momentum injection rate into the ISM via stellar and supernovae feedback, which over time offsets turbulence dissipation and radiative cooling and prevents the ISM from collapsing in the galactic gravitational potential \citep[see also][who propose that feedback maintains the ISM in a marginally Toomre-stable state]{Thompson_etal_2005}.
The local SFR required to keep the ISM in a long-term equilibrium is thus ultimately set by the weight of the ISM in the galactic potential.
Since we expect the ISM in massive, star-forming disk galaxies to exist in such a thermal and dynamical equilibrium, it implies a proportionality between $\SigSFR$ and the weight of the ISM per unit area, commonly referred to as the dynamical equilibrium pressure, $\PDE$.
Their ratio $\Yfb=\PDE/\SigSFR$ is named the feedback yield, as it quantifies the ISM pressure resulting from the injection of momentum and energy by feedback, measured per unit stellar mass formed \citep[see][for a recent formulation of this theory]{Ostriker_Kim_2022}.

Figure~\ref{fig:PR} shows the $\SigSFR$--$\PDE$ relation, hereafter referred to as the pressure-regulated SF relation (PR relation), and the distribution of the $\Yfb$ parameter.
This analysis covers a subsample of 48 galaxies that have \HI\ 21~cm data available to us (see Section~\ref{sec:data} and \citetalias{Sun_etal_2022}).
The measurements shown here represent a major update over those presented in \citetalias{Sun_etal_2020b}, which only covered 28 galaxies and relied on earlier versions of the PHANGS--ALMA CO data and associated \HI\ data.

We fit a power law model to the PR relation with the following parametrization:
\begin{align}
\log_{10}&\!\left(\frac{\SigSFR}{\uSigSFR}\right)\nonumber\\
&= \alpha+\beta\log_{10}\!\left(\frac{\PDE}{10^4\,k\sbsc{B}\,\uP}\right)~.
\label{eq:PR}
\end{align}
Across our sample, the PR relation exhibits a mildly sub-linear slope for several methodological choices (including the fiducial), which is broadly consistent with the results reported in \citetalias{Sun_etal_2020b} and in other studies \citep[e.g.,][]{Fisher_etal_2019,Fisher_etal_2022,Barrera-Ballesteros_etal_2021a}.
When adopting the \citetalias{Bolatto_etal_2013} or \citetalias{Gong_etal_2020} $\alphaCO$, the slope appears near-unity, although still not as steep as that seen in numerical simulations \citep[$\beta$=1.1--1.2; see][]{Kim_etal_2013,Ostriker_Kim_2022}.
This small discrepancy may be due to a limited range of conditions ($\SigSFR\approx10^{-3}$--$10^{-1}\,\uSigSFR$) probed in the observations.
Alternatively, it could be attributed to (1) an elevated $\Yfb$ in high $\SigSFR$ regions, possibly caused by more efficient feedback from clustered supernovae in reality \citep[e.g.,][]{Gentry_etal_2019,Fisher_etal_2019,Fisher_etal_2022}, or (2) additional sources of turbulence injection, such as gas radial inflows \citep[e.g.,][]{Krumholz_etal_2018,Girard_etal_2021}.
Otherwise, the observed $\Yfb$ range of 1--3$\times10^3\,\uV$ is broadly in line with simulation results \citep[e.g.,][]{Kim_etal_2013,Kim_etal_2017,Keller_etal_2022,Ostriker_Kim_2022}.

The intrinsic scatter around the PR relation is mildly smaller than the mKS relation and more comparable to the other two relations (Table~\ref{tab:fit}).
This means the PR relation makes as good empirical predictions for $\SigSFR$ as the mES and FFTR relations, but with similar limitations given its possibly sub-linear slope and thus a varying $\Yfb$.
Practical applications of the PR relation for this purpose should also consider the systematic trends in $\Yfb$ with $\SigSFR$ and other environmental conditions.

We note that among the four SF laws examined in this work, only the PR relation includes both the molecular and the atomic gas.
The first three relations focus only on the molecular gas, which means that they implicitly take the atomic-to-molecular phase balance as given and do not capture any physics related to that process.
In contrast, the PR relation \textit{has} to include both phases because it concerns the energy and momentum budget, which has no natural border between the phases.
To some degree, this makes the PR relation more generally applicable, even to regions or galaxies with no detectable molecular gas \citep[e.g.,][]{Kado-Fong_etal_2022b}.
Nevertheless, the need for \HI\ data limits our ability to study the PR relation for a larger fraction of the PHANGS-ALMA sample.
This situation will likely improve as we gather more \HI\ data for PHANGS targets with VLA and MeerKAT (A.~Sardone et al.\ in preparation; C.~Eibensteiner et al.\ in preparation).

\subsection{Impacts of Methodological Choices}
\label{sec:results:methods}

For all four SF laws, we observe coherent, systematic changes in their shapes when we adopt different methodological choices (Table~\ref{tab:fit}).
Changes in the best-fit power law parameters due to methodological choices are much larger than their formal statistical uncertainties, which suggests that methodology-related systematics are clearly a dominant source of uncertainties.
Here we briefly summarize these systematic changes and discuss their implications.
Note that while the number of available measurements differs for each choice (Table~\ref{tab:fit}), the trends we see remain the same when doing rigorous comparisons with a matched sample of measurements\footnote{We do not show the quantitative results here for the sake of brevity, but note that these results can be easily reproduced with the published data products described in Appendix~\ref{apdx:mega-table}.}.

Considering the impacts of both SFR calibrations and $\alphaCO$ prescriptions, we see that the former can introduce at least a 10--15\% change in the slopes and a 0.15--0.25~dex variation in the normalization factors of the SF laws\footnote{These likely represent only lower limits because the three SFR prescriptions considered here were calibrated in the same way \citep{Belfiore_etal_2023} and thus have some built-in uniformity.}.
The latter can produce a 20--25\% difference in the slopes and 0.10--0.20~dex in the normalization.
All these systematics reflect real uncertainties in our empirical knowledge of ISM physics (such as heating and shielding) and its interactions with stellar populations.

Between the three SFR calibrations, we find that the FUV+22$\mu$m calibration yields slightly higher $\SigSFR$ values at the low end compared to the fiducial H$\alpha$+22$\mu$m calibration, resulting in shallower SF law slopes.
This is in line with the findings by \citet{Belfiore_etal_2023} that the former is still somewhat more susceptible to contamination from IR cirrus at $\SigSFR\lesssim10^{-3}\,\uSigSFR$, even though both calibrations have seen substantial improvements in this aspect compared to previous versions \citep{Leroy_etal_2019}.
The MUSE extinction-corrected H$\alpha$ calibration gives more consistent SF law slopes with the fiducial H$\alpha$+22$\mu$m, but it yields a slightly higher SF law normalization (by 0.1--0.2~dex).
This discrepancy can be traced back to an intrinsic zero-point difference between H$\alpha$ maps obtained from MUSE \citep{Emsellem_etal_2022} versus narrow-band observations (A.~Razza et al.\ in preparation), an issue to be addressed in future processing of the narrow-band data.

Between the four $\alphaCO$ prescriptions, we find that the fiducial \citetalias{Sun_etal_2020b} $\alphaCO$ yields higher $\Sigmol$ at the low end than the Galactic $\alphaCO$, as expected from its built-in metallicity dependence.
The \citetalias{Bolatto_etal_2013} prescription produces even higher $\alphaCO$ at the low end due to its stronger, exponential metallicity dependence; it also produces low $\alphaCO$ in high surface density regions, which comes from an extra negative dependence on the total (stellar + gas) surface density that aims to account for elevated gas excitation and velocity dispersion in the ``starburst'' regime.
\citetalias{Gong_etal_2020} gives similarly low $\alphaCO$ at the high surface density end as \citetalias{Bolatto_etal_2013}, which is expected because the extra dependence on CO intensity in the \citetalias{Gong_etal_2020} $\alphaCO$ aims to capture a similar set of physics.
At the low surface density end, the \citetalias{Gong_etal_2020} $\alphaCO$ is closer to the \citetalias{Sun_etal_2020b} $\alphaCO$ as it also features a power law metallicity dependence.
Overall, we expect the \citetalias{Bolatto_etal_2013} and \citetalias{Gong_etal_2020} prescriptions to likely yield more realistic $\alphaCO$ than the other two options given the extra physics they (at least intend to) capture.
That being said, the exact behavior of $\alphaCO$ in the high-density, high-excitation, ``starburst'' regime and the functional form of its metallicity dependence both remain key open questions.

Related to our treatment of $\alphaCO$, our fiducial treatment uses a constant CO~2--1/1--0 ratio of $R\sbsc{21}=0.65$ in combination with the Galactic, \citetalias{Sun_etal_2020b}, and \citetalias{Bolatto_etal_2013} $\alpha_\mathrm{CO\,(1{-}0)}$ values\footnote{The \citetalias{Bolatto_etal_2013} $\alphaCO$ was calibrated primarily with \CO21\ observations assuming a fixed $R_{21}$, so it should not be combined with a varying $R_{21}$.
The \citetalias{Gong_etal_2020} $\alphaCO$ was also explicitly calibrated for the \CO21\ transition and thus does not need an assumed $R\sbsc{21}$.}.
The assumption of a constant $R\sbsc{21}$ is not realistic on its own, especially given recent findings of an approximate scaling of $R\sbsc{21}\propto\SigSFR^{0.15}$ by several studies \citep{Yajima_etal_2020,denBrok_etal_2021,Leroy_etal_2022,Leroy_etal_2023a}.
If we combine such a $\SigSFR$-dependent $R\sbsc{21}$ value with the fiducial $\alpha_\mathrm{CO\,(1{-}0)}$, the SF law slope would increase from $\beta=1.00$ to $\beta/(1-0.15\beta)\approx1.18$, thus agreeing better with the results found when using the \citetalias{Bolatto_etal_2013} or \citetalias{Gong_etal_2020} $\alphaCO$.
While further investigations on $R_{21}$ is beyond the scope of this work, we expect to improve our fiducial $\alphaCO$ treatment in the near future by explicitly incorporating the $R\sbsc{21}$ prescription suggested in \citet{Leroy_etal_2022,Leroy_etal_2023a}.

\begin{deluxetable}{lcccc}
\tablecaption{Best-fit Parameters for the SF Laws\label{tab:fit}}
\tablewidth{0pt}
\tablehead{
\colhead{Method\phs\phs\phs\phs\phs\phs} & 
\colhead{$\alpha$} &
\colhead{$\beta$} &
\colhead{$\sigma$} &
\colhead{$N\sbsc{det}/N\sbsc{uplim}$} \\[-1.5em]
}
\startdata
\\[-1.5em]
\multicolumn{5}{c}{Molecular Kennicutt-Schmidt relation (mKS; Section~\ref{sec:results:mKS})} \\
\hline
Fiducial & -2.40 & 1.00 & 0.36 & 1585/450 \\
FUV+W4 SFR & -2.34 & 0.88 & 0.29 & 2279/158 \\
Av-corr H$\alpha$ SFR & -2.23 & 0.93 & 0.29 & 974/0 \\
MW $\alpha_\mathrm{CO}$ & -2.43 & 0.92 & 0.37 & 1553/451 \\
B13 $\alpha_\mathrm{CO}$ & -2.36 & 1.21 & 0.38 & 1016/225 \\
G20 $\alpha_\mathrm{CO}$ & -2.22 & 1.18 & 0.35 & 1298/225 \\
\hline
\multicolumn{5}{c}{Molecular Elmegreen--Silk relation (mES; Section~\ref{sec:results:mES})} \\
\hline
Fiducial & -2.23 & 0.77 & 0.31 & 1001/169 \\
FUV+W4 SFR & -2.19 & 0.67 & 0.26 & 1198/31 \\
Av-corr H$\alpha$ SFR & -2.06 & 0.78 & 0.28 & 516/0 \\
MW $\alpha_\mathrm{CO}$ & -2.26 & 0.69 & 0.33 & 999/167 \\
B13 $\alpha_\mathrm{CO}$ & -2.17 & 0.90 & 0.32 & 666/97 \\
G20 $\alpha_\mathrm{CO}$ & -2.11 & 0.77 & 0.33 & 908/121 \\
\hline
\multicolumn{5}{c}{Free-fall time regulated SF relation (FFTR; Section~\ref{sec:results:FFTR})} \\
\hline
Fiducial & -2.32 & 0.65 & 0.34 & 1457/311 \\
FUV+W4 SFR & -2.28 & 0.57 & 0.28 & 2036/66 \\
Av-corr H$\alpha$ SFR & -2.16 & 0.62 & 0.28 & 880/0 \\
MW $\alpha_\mathrm{CO}$ & -2.34 & 0.62 & 0.34 & 1439/314 \\
B13 $\alpha_\mathrm{CO}$ & -2.29 & 0.75 & 0.36 & 970/178 \\
G20 $\alpha_\mathrm{CO}$ & -2.20 & 0.76 & 0.33 & 1294/220 \\
\hline
\multicolumn{5}{c}{Pressure regulated SF relation (PR; Section~\ref{sec:results:PR})} \\
\hline
Fiducial & -2.95 & 0.93 & 0.33 & 1138/313 \\
FUV+W4 SFR & -2.84 & 0.84 & 0.24 & 1696/133 \\
Av-corr H$\alpha$ SFR & -2.72 & 0.85 & 0.25 & 651/0 \\
MW $\alpha_\mathrm{CO}$ & -2.94 & 0.86 & 0.33 & 1122/309 \\
B13 $\alpha_\mathrm{CO}$ & -2.95 & 1.08 & 0.32 & 1015/224 \\
G20 $\alpha_\mathrm{CO}$ & -2.87 & 1.05 & 0.31 & 952/171 \\
\enddata
\tablecomments{
The values reported here are: power law normalization ($\alpha$, in dex units), slope ($\beta$), intrinsic scatter ($\sigma$, in dex units), and the number of detections and upper limits used in the model fit ($N\sbsc{det}/N\sbsc{uplim}$).
See Equations~\ref{eq:mKS}--\ref{eq:PR} for the exact parametrization for each SF law.
The formal statistical uncertainties on the best-fit parameters are not listed here as they are negligible compared to the systematic uncertainties associated with SFR calibrations and $\alphaCO$ prescriptions.
}
\vspace{-2.1\baselineskip}
\end{deluxetable}
\vspace{-1\baselineskip}


\section{Conclusions} \label{sec:conclusion}

In this Letter, we examine four ``star formation laws'' commonly used in the literature based on resolved, multiwavelength measurements of SFR and ISM properties across 80 nearby galaxies.
This work represents a major improvement over several previous studies \citep[e.g.,][]{Wong_Blitz_2002,Bigiel_etal_2011,Leroy_etal_2013a,Utomo_etal_2018,Sun_etal_2020b,Querejeta_etal_2021} given the larger number of galaxies covered, the higher quality of the underlying observational data, and the consistent methodological treatments for deriving relevant physical quantities (especially SFR and molecular gas mass).

We measure the slopes, normalization factors, and intrinsic scatters of the four SF laws as well as their dependence on methodological choices.
We also report the distributions of the ``proportionality constants'' for the SF laws, which quantify star formation efficiencies and feedback yield.
Our key findings are as follows:
\begin{enumerate}

\item Within the range of conditions probed in our sample, each of the four SF laws is well described by a single power law, with typical intrinsic scatter of 0.3--0.4~dex.
For any given set of methodological choices, the molecular Kennicutt-Schmidt relation consistently shows a $\sim$10\% larger intrinsic scatter than the other three SF laws.
The latter relations can thus provide slightly better empirical predictions for the local SFR surface density.

\item Modulo systematic uncertainties related to methodological choices, we find a near-unity slope ($\beta\approx0.9{-}1.2$) for the molecular Kennicutt-Schmidt relation, which implies a roughly constant molecular gas depletion time of 1--3~Gyr.
The molecular Elmegreen--Silk relation and the free-fall time regulated SF relation both have sub-linear slopes ($\beta\approx0.6{-}0.9$) for most methodological choices, which means that the SFE per orbital time (typically 5--10\%) and the SFE per free-fall time (typically 0.5--1\%) both become lower under higher surface density conditions.
The pressure regulated SF relation is also mildly sub-linear in most cases ($\beta\approx0.8{-}1.0$), signifying a potential increase in the feedback yield (typically $1\text{--}3\times10^3\;\uV$) or possible contributions from other turbulence driving mechanisms in high surface density environments.

\item The exact shapes of the SF laws and the distribution functions of physical parameters vary systematically with the adopted SFR calibration and CO-to-H$_2$ conversion factor.
The former introduces at least a 10--15\% uncertainty on the SF law slopes and 0.15--0.25~dex on the normalization, whereas the latter produces differences of 20--25\% for the slopes and 0.10--0.20~dex for the normalization.
This is a general issue applicable to not only the SFR calibrations and conversion factors examined in this work, but also others used in the literature.
It remains the dominant sources of uncertainties for observational studies of SF laws.
Comparisons between different observational datasets as well as theoretical interpretations hinging heavily on the SF law slopes should be particularly cautious about these systematics.

\end{enumerate}

Looking forward, we expect studies that probe more extreme environmental conditions (e.g., very high or very low surface densities) to reveal possible changes of behavior in any of the SF laws, which would indicate additional physics at play.
With much observational efforts already devoted to these directions \citep[e.g.,][]{Daddi_etal_2010,Genzel_etal_2010,delosReyes_Kennicutt_2019,Wilson_etal_2019,Kennicutt_DeLosReyes_2021,Fisher_etal_2022,Kado-Fong_etal_2022b}, the next critical step would be to build a large, homogeneous dataset, to which one can apply improved, consistent methodological treatments.
To make the most of these measurements, it will also be critical to compare the systematic trends that we do observe in the SFE per orbital time, SFE per free-fall time, and feedback yield, to the results of current analytical and numerical models, so as to understand how well these models can describe real star-forming galaxies \citep[e.g.,][S.~Meidt et al., in preparation]{Ostriker_Kim_2022}.
Finally, with reliable measurements of the molecular disk scale height in PHANGS galaxies \citep[J.~Sun et al.\ in preparation]{Jeffreson_etal_2022}, we will expand our analysis in this work to cover the volumetric SF laws originally suggested by \citet{Schmidt_1959} and actively reconsidered in recent works \citep[e.g.,][]{Bacchini_etal_2019a}.


\vspace{\baselineskip}
{
This work was carried out as part of the PHANGS collaboration.
We thank the anonymous referee for constructive feedback and S.~Ellison for helpful discussions.

JS acknowledges support by the Natural Sciences and Engineering Research Council of Canada (NSERC) through a Canadian Institute for Theoretical Astrophysics (CITA) National Fellowship.
The work of AKL is partially supported by the National Science Foundation (NSF) under Grants No.~1615105, 1615109, and 1653300.
The work of ECO was partly supported by grant No.~510940 from the Simons Foundation.
ER acknowledges the support of NSERC, funding reference number RGPIN-2022-03499.
The research of CDW is supported by grants from NSERC and the Canada Research Chairs program.
GAB acknowledges the support from ANID Basal project FB210003.
AH was supported by the Programme National Cosmology et Galaxies (PNCG) of CNRS/INSU with INP and IN2P3, co-funded by CEA and CNES, and by the Programme National ``Physique et Chimie du Milieu Interstellaire'' (PCMI) of CNRS/INSU with INC/INP co-funded by CEA and CNES.
EWK acknowledges support from the Smithsonian Institution as a Submillimeter Array (SMA) Fellow and the Natural Sciences and Engineering Research Council of Canada.
KK gratefully acknowledges funding from the Deutsche Forschungsgemeinschaft (DFG) in the form of an Emmy Noether Research Group (grant number KR4598/2-1, PI Kreckel). 
HAP acknowledges support by the National Science and Technology Council of Taiwan under grant 110-2112-M-032-020-MY3.
JP acknowledges support by the DAOISM grant ANR-21-CE31-0010 and by the Programme National ``PCMI'' of CNRS/INSU with INC/INP, co-funded by CEA and CNES.
MQ acknowledges support from the Spanish grant PID2019-106027GA-C44, funded by MCIN/AEI/10.13039/501100011033.
AS is supported by an NSF Astronomy and Astrophysics Postdoctoral Fellowship under award AST-1903834.
AU acknowledges support from the Spanish grants PGC2018-094671-B-I00, funded by MCIN/AEI/10.13039/501100011033 and by ``ERDF A way of making Europe'', and PID2019-108765GB-I00, funded by MCIN/AEI/10.13039/501100011033.
ES and TGW acknowledge funding from the European Research Council (ERC) under the European Union’s Horizon 2020 research and innovation programme (grant agreement No.~694343).
FBi acknowledges funding from the ERC under the European Union’s Horizon 2020 research and innovation programme (grant agreement No.~726384/Empire).
ADB acknowledges partial support from NSF-AST2108140.
MC gratefully acknowledges funding from the DFG through an Emmy Noether Research Group (grant number CH2137/1-1). COOL Research DAO is a Decentralized Autonomous Organization supporting research in astrophysics aimed at uncovering our cosmic origins.
JG gratefully acknowledges financial support from the Swiss National Science Foundation (grant no CRSII5\_193826).
SCOG and RSK acknowledge funding from the ERC via the Synergy Grant ``ECOGAL'' (project ID~855130). They also acknowledge funding from the DFG via the Collaborative Research Center (SFB 881 -- 138713538) ``The Milky Way System'' (subprojects A1, B1, B2 and B8), from the Heidelberg Cluster of Excellence (EXC 2181 - 390900948) ``STRUCTURES'', and from the German Ministry for Economic Affairs and Climate Action in project ``MAINN'' (funding ID~50OO2206).
KG is supported by the Australian Research Council through the Discovery Early Career Researcher Award (DECRA) Fellowship DE220100766 funded by the Australian Government. 
KG is supported by the Australian Research Council Centre of Excellence for All Sky Astrophysics in 3 Dimensions (ASTRO~3D), through project number CE170100013.
JMDK gratefully acknowledges funding from the ERC under the European Union's Horizon 2020 research and innovation programme via the ERC Starting Grant MUSTANG (grant number 714907). 

This paper makes use of the following ALMA data, which have been processed as part of the PHANGS--ALMA \CO21 survey: \\
\noindent ADS/JAO.ALMA\#2012.1.00650.S, \linebreak 
ADS/JAO.ALMA\#2013.1.00803.S, \linebreak 
ADS/JAO.ALMA\#2013.1.01161.S, \linebreak 
ADS/JAO.ALMA\#2015.1.00121.S, \linebreak 
ADS/JAO.ALMA\#2015.1.00782.S, \linebreak 
ADS/JAO.ALMA\#2015.1.00925.S, \linebreak 
ADS/JAO.ALMA\#2015.1.00956.S, \linebreak 
ADS/JAO.ALMA\#2016.1.00386.S, \linebreak 
ADS/JAO.ALMA\#2017.1.00392.S, \linebreak 
ADS/JAO.ALMA\#2017.1.00766.S, \linebreak 
ADS/JAO.ALMA\#2017.1.00886.L, \linebreak 
ADS/JAO.ALMA\#2018.1.01321.S, \linebreak 
ADS/JAO.ALMA\#2018.1.01651.S, \linebreak 
ADS/JAO.ALMA\#2018.A.00062.S. \linebreak 
ALMA is a partnership of ESO (representing its member states), NSF (USA), and NINS (Japan), together with NRC (Canada), NSC and ASIAA (Taiwan), and KASI (Republic of Korea), in cooperation with the Republic of Chile. The Joint ALMA Observatory is operated by ESO, AUI/NRAO, and NAOJ. The National Radio Astronomy Observatory (NRAO) is a facility of NSF operated under cooperative agreement by Associated Universities, Inc (AUI).

This work is based in part on observations made with NSF's Karl~G.~Jansky Very Large Array 
(VLA; project code:
14A-468, 14B-396, 16A-275, 17A-073, 
18B-184). 
VLA is also operated by NRAO.

This work is based in part on observations made with the Australia Telescope Compact Array (ATCA). ATCA is part of the Australia Telescope National Facility, which is funded by the Australian Government for operation as a National Facility managed by CSIRO.

This work is based in part on observations made with the Westerbork Synthesis Radio Telescope (WSRT) owned by ASTRON. ASTRON, the Netherlands Institute for Radio Astronomy, is an institute of the Dutch Research Council (De Nederlandse Organisatie voor Wetenschappelijk Onderzoek, NWO).

This work is based in part on observations collected at the European Southern Observatory under ESO programmes~094.C-0623 (PI: Kreckel), 095.C-0473,  098.C-0484 (PI: Blanc), 1100.B-0651 (PHANGS-MUSE; PI: Schinnerer), as well as 094.B-0321 (MAGNUM; PI: Marconi), 099.B-0242, 0100.B-0116, 098.B-0551 (MAD; PI: Carollo) and 097.B-0640 (TIMER; PI: Gadotti).

This work is based in part on observations made with the \textit{Spitzer Space Telescope}, which is operated by the Jet Propulsion Laboratory, California Institute of Technology under a contract with NASA.

This work makes use of data products from the \textit{Wide-field Infrared Survey Explorer (WISE)}, which is a joint project of the University of California, Los Angeles, and the Jet Propulsion Laboratory/California Institute of Technology, funded by NASA.

This work is based in part on observations made with the \textit{Galaxy Evolution Explorer (GALEX)}. \textit{GALEX} is a NASA Small Explorer, whose mission was developed in cooperation with the Centre National d'Etudes Spatiales (CNES) of France and the Korean Ministry of Science and Technology. \textit{GALEX} is operated for NASA by the California Institute of Technology under NASA contract NAS5-98034.

This work is based in part on data gathered with the CIS 2.5m Ir\'en\'ee du Pont Telescope and the ESO/MPG 2.2m Telescope at Las Campanas Observatory, Chile.

This work has made use of the NASA/IPAC Infrared Science Archive (IRSA) and the NASA/IPAC Extragalactic Database (NED), which are funded by NASA and operated by the California Institute of Technology.

We acknowledge the usage of the SAO/NASA Astrophysics Data System.

\facilities{
ALMA, VLA, ATCA, WSRT, VLT:Yepun, Spitzer, WISE, GALEX, Du~Pont, Max~Planck:2.2m, IRSA
}

\software{
\texttt{NumPy} \citep{NumPy_2020},
\texttt{Astropy} \citep{Astropy_2013,Astropy_2018},
\texttt{Matplotlib} \citep{Matplotlib_2007},
\texttt{linmix} \citep{Kelly_2007},
\texttt{MegaTable} \citep{MegaTable_3.0},
\texttt{adstex} (\url{https://github.com/yymao/adstex}).
}

}


\appendix
\restartappendixnumbering

\section{Data Products}
\label{apdx:mega-table}

This paper uses the latest version of the PHANGS high-level measurement database (\citetalias{Sun_etal_2022}).
This evolving database incorporates homogenized measurements from a rich set of multiwavelength observations and enables rigorous comparisons across all galaxies.
\citetalias{Sun_etal_2022} already presented the database construction methodologies (such as observational data sampling and weighing, conversion to physical quantities) in great detail.
Here we briefly describe the major improvements we have incorporated since the first publication of the database (v3.0) with \citetalias{Sun_etal_2022}.

A first major improvement is the ingestion of data products from the PHANGS--MUSE observations \citep{Emsellem_etal_2022}.
In particular, the analysis in this paper involves the H$\alpha$ line intensity maps and the associated uncertainty maps, which are part of the PHANGS--MUSE Data Analysis Pipeline high-level products \citep[or DAP products; see Section~5 in][]{Emsellem_etal_2022}.
We correct these maps for dust extinction pixel-by-pixel based on the observed intensity ratio of H$\alpha$ and H$\beta$ (which is also part of the DAP products).
We then calculate the average H$\alpha$ surface brightness over each 1.5~kpc region from both the uncorrected and corrected H$\alpha$ maps.
These average surface brightness values, along with their uncertainties (from Gaussian error propagation), are recorded in our high-level database and used to further derive SFR surface densities and their uncertainties (see Section~\ref{sec:data}).

A second improvement is on the \HI\ 21~cm line data used for measuring the atomic gas surface density.
The older v3.0 database already incorporated \HI\ data from a variety of surveys (such as VLA:THINGS--\citealt{Walter_etal_2008}, VLA:VIVA--\citealt{Chung_etal_2009}, ATCA:LVHIS--\citealt{Koribalski_etal_2018}, and VLA:PHANGS--A.~Sardone et al.\ in preparation), and all \HI\ measurements were derived from the ``official'' moment maps produced by each of the survey teams.
However, each team created these moment maps from the original data cubes in a slightly different way, which led to an extra layer of inhomogeneity among these data products.
We have now reprocessed all \HI\ data starting from their raw data cubes, adopting a uniform set of treatments for cube post-processing, signal masking, and moment map generation across all surveys.
We have also processed additional \HI\ data from the HALOGAS \citep{Heald_etal_2011} and WHISP \citep{vanderHulst_etal_2001} surveys and added them into the mix.
These improvements allow us to include \HI\ measurements for more galaxies and allow fairer comparisons between galaxies covered by different surveys.

A third improvement is on the data averaging scheme when extracting measurements for each 1.5~kpc area.
For the older v3.0 database, we extracted ``area-weighted'' average values by directly averaging over all pixels in a native resolution image that fall inside the (sharp) boundary of each 1.5~kpc aperture.
This approach was well suited for the analyses in \citetalias{Sun_etal_2022}, which required maximally independent measurements between adjacent apertures and consistent averaging schemes between cloud-scale molecular gas measurements and large-scale environmental measurements (see Section~3 therein).
However, this approach can lead to slightly different ``effective smoothing scales'' among datasets and galaxies, as the image native resolution vary from case to case even though the averaging aperture size is fixed.
Such behavior is not ideal for the analyses in this work, because the SF laws are known to vary as a function of spatial scales \citep[e.g.,][]{Kreckel_etal_2018,Pessa_etal_2021}.
Consequently, for all ``aperture-scale averaged'' measurements used in this work, we instead calculate them by first convolving the native resolution images to a matched Gaussian beam with a full-width-half-maximum of 1.5~kpc, and then directly extracting values from the convolved images at the center of each aperture.
This alternative averaging scheme ensures all important quantities used in this work (such as $\SigSFR$, $\Sigmol$, and $\PDE$) are measured on a strictly fixed spatial scale of 1.5~kpc.

To make the improved data products available to the community, we distill all measurements used in this work into a machine-readable table (see Table~\ref{tab:mrt}) and publish it with this paper.
We will also release an associated new version (v4.0) of the full PHANGS high-level database at the \href{https://www.canfar.net/storage/vault/list/phangs/RELEASES/Sun_etal_2022}{same online location} as the version published in \citetalias{Sun_etal_2022}.

\begin{deluxetable}{lll}
\tabletypesize{\footnotesize}
\tablewidth{0pt}
\tablecaption{Column Descriptions for the Machine-Readable Table\label{tab:mrt}}
\tablehead{
\colhead{Column Name} & \colhead{\edit1{Unit}} & \colhead{Description}
}
\startdata
\texttt{\detokenize{gal_name}} &  & Galaxy name \\
\texttt{\detokenize{RA}} & $\mathrm{{}^{\circ}}$ & Right Ascension of the aperture center \\
\texttt{\detokenize{DEC}} & $\mathrm{{}^{\circ}}$ & Declination of the aperture center \\
\texttt{\detokenize{Sigma_SFR_HaW4recal}} & $\mathrm{M_{\odot}\,yr^{-1}\,kpc^{-2}}$ & SFR surface density (H$\alpha$+22\micron) \\
\texttt{\detokenize{e_Sigma_SFR_HaW4recal}} & $\mathrm{M_{\odot}\,yr^{-1}\,kpc^{-2}}$ & Error on SFR surface density (H$\alpha$+22\micron) \\
\texttt{\detokenize{Sigma_SFR_FUVW4recal}} & $\mathrm{M_{\odot}\,yr^{-1}\,kpc^{-2}}$ & SFR surface density (FUV+22\micron) \\
\texttt{\detokenize{e_Sigma_SFR_FUVW4recal}} & $\mathrm{M_{\odot}\,yr^{-1}\,kpc^{-2}}$ & Error on SFR surface density (FUV+22\micron) \\
\texttt{\detokenize{Sigma_SFR_Hacorr}} & $\mathrm{M_{\odot}\,yr^{-1}\,kpc^{-2}}$ & SFR surface density ($A_V$-corrected H$\alpha$) \\
\texttt{\detokenize{e_Sigma_SFR_Hacorr}} & $\mathrm{M_{\odot}\,yr^{-1}\,kpc^{-2}}$ & Error on SFR surface density ($A_V$-corrected H$\alpha$) \\
\texttt{\detokenize{Sigma_mol_S20}} & $\mathrm{M_{\odot}\,pc^{-2}}$ & Molecular gas surface density (S20 $\alphaCO$) \\
\texttt{\detokenize{e_Sigma_mol_S20}} & $\mathrm{M_{\odot}\,pc^{-2}}$ & Error on molecular gas surface density (S20 $\alphaCO$) \\
\texttt{\detokenize{Sigma_mol_MW}} & $\mathrm{M_{\odot}\,pc^{-2}}$ & Molecular gas surface density (Galactic $\alphaCO$) \\
\texttt{\detokenize{e_Sigma_mol_MW}} & $\mathrm{M_{\odot}\,pc^{-2}}$ & Error on molecular gas surface density (Galactic $\alphaCO$) \\
\texttt{\detokenize{Sigma_mol_B13}} & $\mathrm{M_{\odot}\,pc^{-2}}$ & Molecular gas surface density (B13 $\alphaCO$) \\
\texttt{\detokenize{e_Sigma_mol_B13}} & $\mathrm{M_{\odot}\,pc^{-2}}$ & Error on molecular gas surface density (B13 $\alphaCO$) \\
\texttt{\detokenize{Sigma_mol_G20}} & $\mathrm{M_{\odot}\,pc^{-2}}$ & Molecular gas surface density (G20 $\alphaCO$) \\
\texttt{\detokenize{e_Sigma_mol_G20}} & $\mathrm{M_{\odot}\,pc^{-2}}$ & Error on molecular gas surface density (G20 $\alphaCO$) \\
\texttt{\detokenize{Sigma_mol_per_t_orb_S20}} & $\mathrm{M_{\odot}\,yr^{-1}\,kpc^{-2}}$ & Molecular gas surface density per orbital time (S20 $\alphaCO$) \\
\texttt{\detokenize{e_Sigma_mol_per_t_orb_S20}} & $\mathrm{M_{\odot}\,yr^{-1}\,kpc^{-2}}$ & Error on molecular gas surface density per orbital time (S20 $\alphaCO$) \\
\texttt{\detokenize{Sigma_mol_per_t_orb_MW}} & $\mathrm{M_{\odot}\,yr^{-1}\,kpc^{-2}}$ & Molecular gas surface density per orbital time (Galactic $\alphaCO$) \\
\texttt{\detokenize{e_Sigma_mol_per_t_orb_MW}} & $\mathrm{M_{\odot}\,yr^{-1}\,kpc^{-2}}$ & Error on molecular gas surface density per orbital time (Galactic $\alphaCO$) \\
\texttt{\detokenize{Sigma_mol_per_t_orb_B13}} & $\mathrm{M_{\odot}\,yr^{-1}\,kpc^{-2}}$ & Molecular gas surface density per orbital time (B13 $\alphaCO$) \\
\texttt{\detokenize{e_Sigma_mol_per_t_orb_B13}} & $\mathrm{M_{\odot}\,yr^{-1}\,kpc^{-2}}$ & Error on molecular gas surface density per orbital time (B13 $\alphaCO$) \\
\texttt{\detokenize{Sigma_mol_per_t_orb_G20}} & $\mathrm{M_{\odot}\,yr^{-1}\,kpc^{-2}}$ & Molecular gas surface density per orbital time (G20 $\alphaCO$) \\
\texttt{\detokenize{e_Sigma_mol_per_t_orb_G20}} & $\mathrm{M_{\odot}\,yr^{-1}\,kpc^{-2}}$ & Error on molecular gas surface density per orbital time (G20 $\alphaCO$) \\
\texttt{\detokenize{Sigma_mol_per_t_ff_S20}} & $\mathrm{M_{\odot}\,yr^{-1}\,kpc^{-2}}$ & Molecular gas surface density per free-fall time (S20 $\alphaCO$) \\
\texttt{\detokenize{e_Sigma_mol_per_t_ff_S20}} & $\mathrm{M_{\odot}\,yr^{-1}\,kpc^{-2}}$ & Error on molecular gas surface density per free-fall time (S20 $\alphaCO$) \\
\texttt{\detokenize{Sigma_mol_per_t_ff_MW}} & $\mathrm{M_{\odot}\,yr^{-1}\,kpc^{-2}}$ & Molecular gas surface density per free-fall time (Galactic $\alphaCO$) \\
\texttt{\detokenize{e_Sigma_mol_per_t_ff_MW}} & $\mathrm{M_{\odot}\,yr^{-1}\,kpc^{-2}}$ & Error on molecular gas surface density per free-fall time (Galactic $\alphaCO$) \\
\texttt{\detokenize{Sigma_mol_per_t_ff_B13}} & $\mathrm{M_{\odot}\,yr^{-1}\,kpc^{-2}}$ & Molecular gas surface density per free-fall time (B13 $\alphaCO$) \\
\texttt{\detokenize{e_Sigma_mol_per_t_ff_B13}} & $\mathrm{M_{\odot}\,yr^{-1}\,kpc^{-2}}$ & Error on molecular gas surface density per free-fall time (B13 $\alphaCO$) \\
\texttt{\detokenize{Sigma_mol_per_t_ff_G20}} & $\mathrm{M_{\odot}\,yr^{-1}\,kpc^{-2}}$ & Molecular gas surface density per free-fall time (G20 $\alphaCO$) \\
\texttt{\detokenize{e_Sigma_mol_per_t_ff_G20}} & $\mathrm{M_{\odot}\,yr^{-1}\,kpc^{-2}}$ & Error on molecular gas surface density per free-fall time (G20 $\alphaCO$) \\
\texttt{\detokenize{P_DE_S20}} & $k_\mathrm{B}\,\mathrm{K\,cm^{-3}}$ & ISM dynamical equilibrium pressure (S20 $\alphaCO$) \\
\texttt{\detokenize{e_P_DE_S20}} & $k_\mathrm{B}\,\mathrm{K\,cm^{-3}}$ & Error on ISM dynamical equilibrium pressure (S20 $\alphaCO$) \\
\texttt{\detokenize{P_DE_MW}} & $k_\mathrm{B}\,\mathrm{K\,cm^{-3}}$ & ISM dynamical equilibrium pressure (Galactic $\alphaCO$) \\
\texttt{\detokenize{e_P_DE_MW}} & $k_\mathrm{B}\,\mathrm{K\,cm^{-3}}$ & Error on ISM dynamical equilibrium pressure (Galactic $\alphaCO$) \\
\texttt{\detokenize{P_DE_B13}} & $k_\mathrm{B}\,\mathrm{K\,cm^{-3}}$ & ISM dynamical equilibrium pressure (B13 $\alphaCO$) \\
\texttt{\detokenize{e_P_DE_B13}} & $k_\mathrm{B}\,\mathrm{K\,cm^{-3}}$ & Error on ISM dynamical equilibrium pressure (B13 $\alphaCO$) \\
\texttt{\detokenize{P_DE_G20}} & $k_\mathrm{B}\,\mathrm{K\,cm^{-3}}$ & ISM dynamical equilibrium pressure (G20 $\alphaCO$) \\
\texttt{\detokenize{e_P_DE_G20}} & $k_\mathrm{B}\,\mathrm{K\,cm^{-3}}$ & Error on ISM dynamical equilibrium pressure (G20 $\alphaCO$) \\
\enddata
\tablecomments{The full machine-readable table is published in its entirety both in the electronic edition of the journal and in the \href{https://www.canfar.net/storage/vault/list/phangs/RELEASES/Sun_etal_2023}{PHANGS CADC repository}. A column-by-column description is shown here for guidance regarding its form and content.}
\end{deluxetable}

\section{Fitting Star Formation Laws with \texttt{linmix}}
\label{apdx:fit}

In Section~\ref{sec:results}, we present the best-fit power law models for each of the four star formation laws measured with each of the six methodological choices.
This appendix describes how we determine the best-fit models, the assumptions involved in the model fit, and (most importantly) our treatments for non-detections.

\begin{figure*}[ht]
\vspace{-1.0\baselineskip}
\gridline{\fig{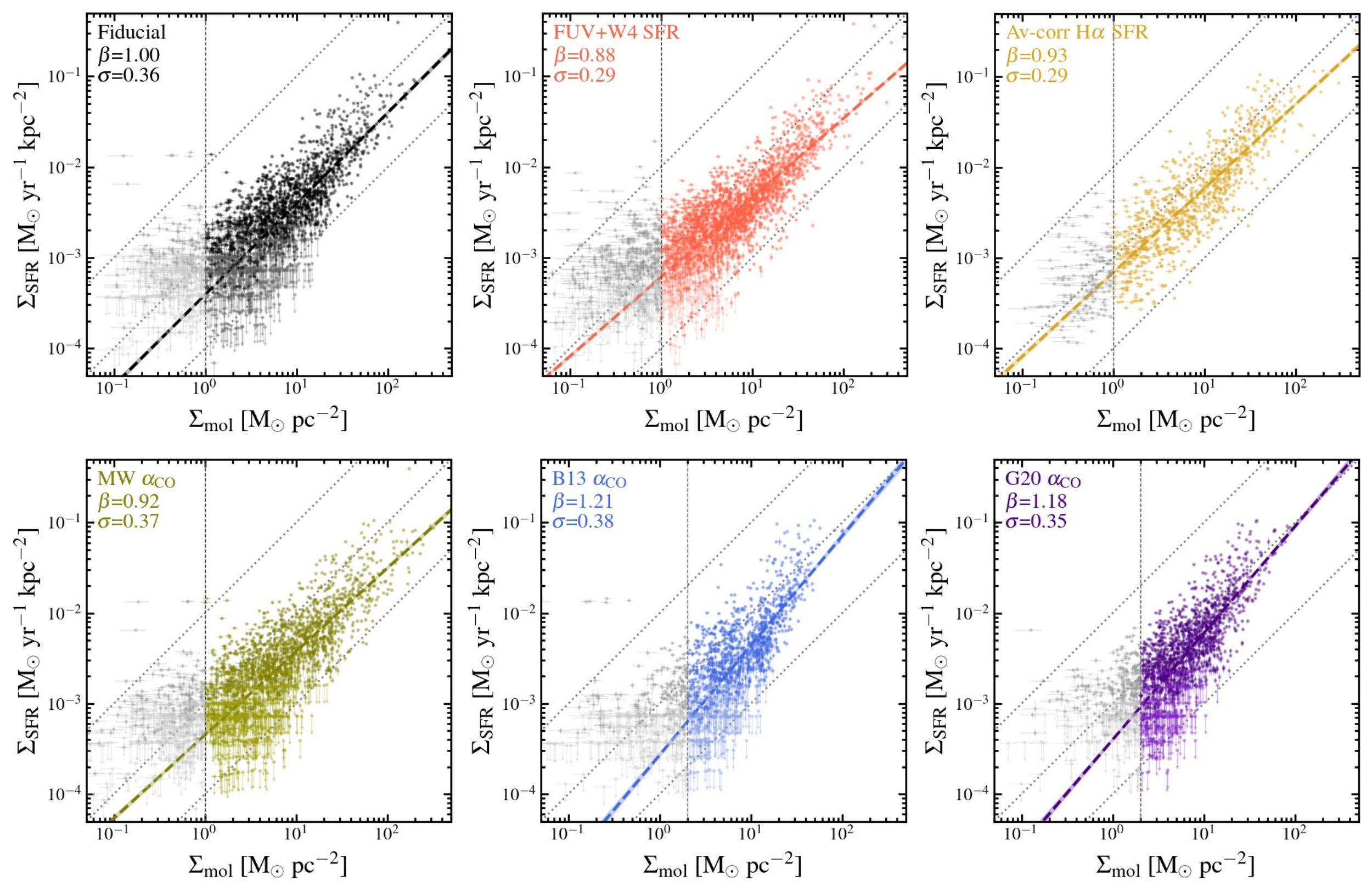}{0.99\textwidth}{}}
\vspace{-2.5\baselineskip}
\caption{
An example figure from the figure set, showing our measurements and best-fit power law models for the molecular Kennicutt-Schmidt relation and how they vary with methodological choices (i.e., SFR calibrations and CO-to-H$_2$ conversion factors).
In each panel, the best-fit power law slope $\beta$ and intrinsic scatter $\sigma$ are displayed at the top left corner;
the vertical dashed line marks the $\Sigmol$ threshold above which the power law model fit is performed ($\sim1\;\uSig$ with the fiducial and MW $\alphaCO$, and $\sim2\;\uSig$ with the other $\alphaCO$).
The complete figure set (4 images) is available in the online journal.
}
\vspace{0.5\baselineskip}
\label{fig:fit}
\end{figure*}

For each star formation law measured with each specific set of methodological choice, we determine a best-fit linear relation in the log--log space, which translates to a power law relation in linear space.
We perform this model fit with the \texttt{linmix} package \citep{Kelly_2007}.
The model fit assumes that the two-dimensional data distribution in logarithmic space can be described by a single, underlying linear relation, with residual scatter attributed to a combination of measurement uncertainties (in both $x$ and $y$) and a fixed intrinsic scatter (along the $y$ direction).
The posterior distributions for all model parameters, including nuisance parameters (e.g., those in a Gaussian mixture model describing the data distribution along the $x$ direction), are evaluated in a hierarchical Bayesian framework with Markov-Chain Monte Carlo.
The best-fit value for each parameter is then determined from the median value over all realizations (see Figure~\ref{fig:fit} for visualizations of the best-fit models).
Since the joint posterior distribution of all parameters very closely resembles an N-dimensional Gaussian, using its maxima would yield almost identical results.

The \texttt{linmix} model can self-consistently handle data censoring (i.e., non-detections) for the dependent variable, but \textit{not} for the independent variable.
However, failure to account for the latter could introduce biases due to increasingly incomplete sampling towards the lower end of the relation.
This is particularly important for the mKS relation, where the independent variable ($\Sigmol$) is translated directly from CO line intensity and thus strongly affected by data sensitivity limit at low $\Sigmol$.
To address this issue, we calculate the data number density per log $\Sigmol$ interval in each $\Sigmol$ bin and determine a $\Sigmol$ threshold below which the data density drops to near half of the maximum value.
This $\Sigmol$ threshold roughly coincides with the highest $\Sigmol$ upper limits in our sample, which confirms that it is where data censoring becomes important.
We thus perform the \texttt{linmix} model fit only on measurements (including detections and upper limits for $\SigSFR$) above this threshold, so that the best-fit parameters can be relatively unaffected.
The $\Sigmol$ threshold changes according to the adopted CO-to-H$_2$ conversion factors (see Figure~\ref{fig:fit} for the exact location of this threshold).
This $\Sigmol$-clipping is applied to the model fits for all four SF laws examined in Section~\ref{sec:results} to ensure self-consistency in this work.

We emphasize that correct treatments of data censoring are crucial for deriving unbiased fit parameters, as have been shown in previous studies \citep[e.g.,][]{Pessa_etal_2021}.
This is especially important when dealing with spatially resolved measurements, for which the number of non-detections can be large and their distribution is concentrated towards the lower end of the probed parameter space.
To illustrate this issue, we perform a test fit for the mKS relation with the fiducial SFR calibration and $\alphaCO$, but without including any $\SigSFR$ upper limits.
This yields a much shallower slope of $\beta=0.80$ compared to $\beta=1.00$ when including the upper limits.
This is expected since most $\SigSFR$ upper limits are distributed near the low $\Sigmol$ end of the mKS relation, without which the average $\SigSFR$ value is biased high at low $\Sigmol$, and thus the power-law slope is biased low.
Beside the treatments of data censoring, the handling of measurement uncertainties and choice of regression methods could also affect the fit result \citep[e.g.,][]{delosReyes_Kennicutt_2019,Tabatabaei_etal_2022}.


\bibliography{main.bib}




\suppressAffiliationsfalse
\allauthors


\end{document}